

\message{For printing the figures, you need to have the
files epsf.tex, 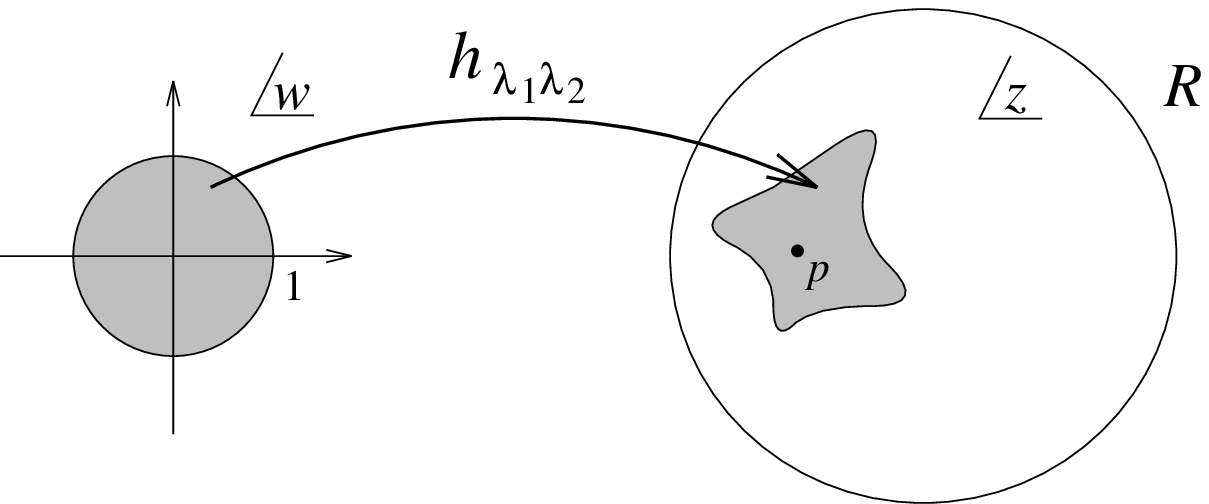 and figure2.eps
in your area. Do you have these files? Enter y/n :    }
\read -1 to \figcount
\if y\figcount \message{This will come out with the figures}
\else \message{This will come out without the figures}\fi

\if y\figcount
    \input epsf
\else\fi

\input phyzzx

\hsize=40pc
%
\catcode`\@=11 
\def\NEWrefmark#1{\step@ver{{\;#1}}}
\catcode`\@=12 

\def\square{\kern1pt\vbox{\hrule height 1.2pt\hbox{\vrule width 1.2pt\hskip 3pt
   \vbox{\vskip 6pt}\hskip 3pt\vrule width 0.6pt}\hrule height 0.6pt}\kern1pt}
\def\d{\ket{D}}
\def\x{\ket{\chi}}

\def\bra#1{\langle #1 |}
\def\ket#1{| #1 \rangle}

\def\ov{{\overline}}

\def\A{{\cal A}}
\def\B{{\cal B}}
\def\C{{\cal C}}

\def\H{\widehat{\cal H}}
\def\HH{{\cal H}}

\def\K{{\cal K}}

\def\L{{\cal L}}
\def\M{{\cal M}}

\def\O{{\cal O}}
\def\P{{\cal P}}

\def\s{{\cal S}}

\def\V{{\cal V}}

\def\p{\partial}

\def\e{{\epsilon}}

\def\k{{\kappa}}

\def\ov{\overline}

\def\wh{\widehat}

\def\B{{\cal B}}

\def\P{{\cal P}}
\def\V{{\cal V}}
\def\O{{\cal O}}
\def\S{{\cal S}}
\def\p{\partial}

\def\bv{{\bf v}}

\def\define#1#2\par{\def#1{\Ref#1{#2}\edef#1{\noexpand\refmark{#1}}}}
\def\con#1#2\noc{\let\?=\Ref\let\<=\refmark\let\Ref=\REFS
         \let\refmark=\undefined#1\let\Ref=\REFSCON#2
         \let\Ref=\?\let\refmark=\<\refsend}

\let\refmark=\NEWrefmark

\define\hata{H. Hata, `Soft dilaton theorem in string field theory',
        Prog. Theor. Phys. {\bf 88}(1992)1197.}

\define\hatanagoshi{H. Hata and Y. Nagoshi, `Dilaton and classical solutions
in pregeometrical string field theory', Prog. Th. Phys. {\bf 80} (1988) 1088.}

\define\yoneya{T. Yoneya, `String-coupling constant and dilaton vacuum
expectation value in string field theory',  Phys. Lett. B {\bf 197} (1987) 76.}

\define\kugozwiebach{T. Kugo and B. Zwiebach, `Target space duality as a
symmetry of string field theory', Prog. Th. Phys. {\bf 87}
                   (1992) 801.}

\define\zwiebachlong{B. Zwiebach, `Closed string field theory: Quantum
action and the Batalin-Vilkovisky master equation', Nucl. Phys {\bf B390}
(1993) 33, hep-th/9206084.}

\define\sonodazwiebach{H. Sonoda and B. Zwiebach, `Closed string field theory
loops with symmetric factorizable quadratic differentials',
Nucl. Phys. {\bf B331} (1990) 592.}

\define\gilkey{P. B. Gilkey, {\it Invariance Theory, the Heat Equation,
and the Atiyah-Singer Index Theorem}, Wilmington, Delaware, Publish
or Perish, 1984.}

\define\nelson{P. Nelson, `Covariant insertion of general vertex operators',
Phys. Rev. Lett. {\bf 62}(1989)993;
H. S. La and P. Nelson, `Effective field equations for fermionic strings',
Nucl. Phys. {\bf B332} (1990) 83.}

\define\distlernelson{
J. Distler and P. Nelson, `Topological couplings and contact terms in
2-D field theory', Comm. Math. Phys. {\bf 138} (1991) 273;
`The Dilaton Equation in Semirigid String Theory',
Nucl. Phys. {\bf B366} (1991) 255.}

\define\polchinski{J. Polchinski, `Factorization of bosonic string amplitudes',
Nucl. Phys. {\bf B307} (1988) 61.}

\define\belopolskyzwiebach{A. Belopolsky and B. Zwiebach,
`Off-shell string amplitudes: Towards a computation
of the tachyon potential', MIT preprint, MIT-CTP-2336, submitted to
Nucl. Phys. B, hep-th/9409015.}

\define\senzwiebach{A. Sen and B. Zwiebach, `Local background independence
of classical closed string field theory', Nucl. Phys. {\bf B414}
(1994) 649,  hep-th/9307088.}

\define\senzwiebachnew{A. Sen and B. Zwiebach, `Background
independent algebraic structures in closed string field theory',
MIT-CTP-2346, August 1994, hep-th/9408053.}

\define\senzwiebachtwo{A. Sen and B. Zwiebach, `Quantum background
independence of closed string field theory', Nucl. Phys. {\bf B423} (1994) 580,
hep-th/9311009.}

\define\campbell{M. Campbell, P. Nelson and E. Wong, Int. Jour. Mod.
Phys. {\bf A6} (1991) 4909.}

\define\ranganathan{K. Ranganathan, `Nearby CFT's in the operator
formalism: the role of a connection', Nucl. Phys. {\bf B408} (1993) 180. }

\define\rangasonodazw{K. Ranganathan, H. Sonoda and B. Zwiebach, `Connections
on the state-space over conformal field theories', Nucl. Phys.
{\bf B414} (1994) 405, hep-th/9304053.}

\define\imbimbo{C. M. Becchi, R. Collina, and C. Imbimbo, `On the semirelative
condition for closed topological strings',
Phys. Lett. B {\bf 322} (1994) 79.}

\define\voronov{T. Kimura, J. Stasheff, and A. A. Voronov, `On operad
structures
of moduli spaces and string theory', Kyoto University preprint RIMS-936,
July 1993, to appear in Comm. Math. Phys., hep-th/9307114.}

\define\voronovkimura{T. Kimura and A. A. Voronov, `The cohomology of algebras
over moduli spaces', University of North Carolina preprint,
October 1994, hep-th/9410108.}

\define\sentalk{A. Sen, `Some applications of string field theory', Proceedings
of the Conference Strings and Symmetries, Stony Brook, May, 1991, World
Scientific, p.355., hep-th/9109022.}

\define\wong{E. Wong, `Recursion Relations in Semirigid
Topological Gravity', Int. Jour. Mod. Phys. {\bf A7} (1992) 6773.}

\define\bankssen{T. Banks, D. Nemeschansky and A. Sen, `Dilaton coupling
and BRST quantization of bosonic strings, Nucl. Phys {\bf B277} (1986) 67. }

\define\tseytlin{E. Fradkin and A. Tseytlin, Phys. Lett. {\bf 158B } (1985)
316;
{\bf 160B} (1985) 64.}

\define\singer{I. M.  Singer, and J. A. Thorpe, {\it  Lecture notes
on elementary topology and geometry}, New York: Springer Verlag (1976).}

\overfullrule=0pt
\baselineskip 16pt plus 1pt minus 1pt
\nopubblock
{}\hfill \vbox{\hbox{MIT-CTP-2376}
\hbox{UFIFT-HEP-94-14}
\hbox{hep-th/9411047} }\break

\title{THE DILATON THEOREM AND CLOSED STRING BACKGROUNDS}

\author{Oren Bergman  \foot{E-mail  address: oren@phys.ufl.edu \hfill\break
Supported in part by D.O.E. contract DE-FG05-86ER-40272.}}
\address{Department of Physics\break
University of Florida\break
Gainesville, FL 32611, U.S.A.}
\author{Barton Zwiebach \foot{E-mail address: zwiebach@irene.mit.edu,
zwiebach@mitlns.mit.edu.\hfill\break Supported in part by D.O.E.
contract DE-AC02-76ER03069.}}
\address{Center for Theoretical Physics,\break
LNS and Department of Physics\break
MIT, Cambridge, Massachusetts 02139, U.S.A.}

\abstract
{The zero-momentum ghost-dilaton is a non-primary BRST physical
state present in every bosonic closed string background.
It is given by the action of the BRST operator on another state
$\x$, but remains nontrivial in the semirelative BRST cohomology.
When local coordinates arise from metrics we show that dilaton
and $\x$ insertions compute Riemannian curvature and geodesic curvature
respectively.
A proper definition of a CFT deformation induced
by the dilaton requires surface integrals of the dilaton and line integrals
of $\x$. Surprisingly, the ghost
number anomaly makes this a trivial deformation. While dilatons cannot
deform conformal theories, they actually deform conformal
string backgrounds, showing in a simple context that
a string background is not necessarily the same as a CFT.
We generalize the earlier proof of quantum background independence of
string theory to show that a dilaton shift amounts to a shift
of the string coupling in the field-dependent part of the quantum string
action. Thus the
``dilaton theorem'', familiar for on-shell string amplitudes,
holds off-shell as a consequence of an exact symmetry
of the string action.}
\endpage

\chapter{Introduction and Summary}

The soft dilaton theorem is a result relating an on-shell string
amplitude with one zero-momentum dilaton and other physical states, to the
on-shell string amplitude of the other physical states without
the dilaton. The simple form that this relation takes
leads to an interesting result.
For the standard bosonic string background
a shift in the vacuum expectation value of the
dilaton field changes the dimensionless string coupling constant $\k$,
and the slope parameter $\alpha'$ (see [\hata]).

Following Yoneya [\yoneya], who first investigated the dilaton
theorem in the context of light-cone string field theory, Hata and Nagoshi
[\hatanagoshi],  Kugo and Zwiebach
[\kugozwiebach], and Hata [\hata], have studied this subject in the context
of covariantized light-cone string field theories. Several conceptual
issues related to operator ordering and the treatment
of zero-momentum dilatons, which necessarily have zero string length, make
it desirable to investigate the dilaton theorem in the context of covariant
closed string field theory.

The dilaton theorem was studied in a more geometrical setting by Distler
and Nelson [\distlernelson], who concentrated on the ghost part $\d$ of the
dilaton state. This BRST invariant physical state exists in {\it every} bosonic
string background. Shifting the vacuum expectation value of the corresponding
space-time field is expected to shift only the dimensionless
coupling constant $\k$.  The dilaton state $\d$ is not primary
and is almost BRST trivial, namely $\d = -Q\x$,
with $\x$ a state that is illegal
since it fails to be annihilated by $b_0^-$ [\distlernelson].
Distler and Nelson considered the correlator of $n$ {\it physical}
states and a dilaton on a genus $g$ Riemann surface $\Sigma$,
with the dilaton insertion integrated over the surface. The aim was
to show that, schematically,
$$ \Bigl\langle \int_{\Sigma}\hskip-3pt d^2z
D(z,\bar z) \,\Phi_1(P_1)\cdots \Phi_n(P_n)
\Bigr\rangle_{\Sigma} \,\,\sim\,
(2-2g-n)\Bigl\langle \Phi_1(P_1)\cdots \Phi_n(P_n)
\Bigr\rangle_{\Sigma} \,. \eqn\distnel$$
This equation states that integrating a dilaton amounts to multiplication
by the Euler number of the surface,
thought of as having $n$ boundaries.
This property, if generalized
to the case when the $n$-insertions and the underlying
moduli of the surface are also integrated, could be used to argue the dilaton
theorem for on-shell string amplitudes.
At first sight Eqn.~\distnel\ looks problematic
because the dilaton, despite being physical,
is not primary and therefore cannot be integrated on a Riemann surface
unless one has fixed local coordinates at every point on the surface.
The integral, a priori, depends on this choice. The main point of
Ref.[\distlernelson] is that while locally the integrand depends on the
choice of coordinates, the integral could capture topological information
that is independent of the choice of coordinates. This expectation
is not fully realized.
As emphasized by the authors, the `$n$' contribution appearing in
the above right hand side arises from ``contact terms'' and is
ambiguous. The difficulty arises when the dilaton approaches any
of the fixed punctures and as a consequence the surface approaches
degeneration.
In order to obtain a
well-defined local coordinate for the dilaton
one must parameterize
the nearly degenerate surfaces.
This can be done by factoring out a three punctured sphere.
This three punctured sphere can be taken to be the standard two-punctured
sphere, i.e.  the $z$-sphere with local coordinates $z_1 = z$ and $z_2 = 1/z$,
with an extra puncture at $z=1$. Upon a
choice of a local coordinate at the extra puncture, one can calculate
unambiguously the so-called contact term. It turns out that
the contact term depends on the choice of this local coordinate, and
therefore the
left hand side of \distnel\  is not well-defined. If the
local coordinate is chosen to be symmetric under the exchange of
the other two punctures the expected result arises
[\distlernelson].\foot{Distler and Nelson argued that the
choice of a symmetric three punctured
sphere was natural from the viewpoint of string field theory, and suggested
further investigation of this issue.}
Our results corroborate that the above left hand side is indeed ambiguous,
and requires a definition.

As the above discussion illustrates, the dilaton theorem
is difficult to establish  within first quantization
(conformal field theory)  due
to the need to deal consistently with degenerate surfaces which must
appear when considering string amplitudes.
These difficulties do not arise in covariant closed string field theory,
where the action does not involve nearly degenerate surfaces.\foot{The Feynman
rules construct the nearly degenerate surfaces
in a way that is consistent with factorization.}
Within string field theory the issue
is to show that a shift of the string field dilaton changes the
string field measure $d\mu e^{2S/\hbar}$ in the same way as a change of
$\k$ would. At the classical level this reduces to the statement that
the change of the classical action upon a shift of the string field
dilaton can be compensated by changing the string field coupling constant.
The aim of the present paper is to prove conclusively the dilaton theorem
using covariant closed string field theory.
This problem is also a problem in background
independence [\senzwiebach, \senzwiebachtwo], and
raises new and subtle complications in the study of this issue.
In this paper we will succeed in
proving the closed string field theory dilaton theorem for the string field
dependent part of the string field measure. The string field
independent terms of the string field measure may or may not work out,
we do not know at present. Such terms required careful treatment
in the proof of background independence [\senzwiebachtwo],
and additional subtleties arise here.

One of the rather interesting surprises in the present analysis
was finding a subtle but
crucial difference between a conformal field theory (CFT) and a
string background.
While a CFT defines
correlators on fixed Riemann surfaces, a string background
includes forms on moduli spaces of Riemann surfaces.
A string background is the proper data for the construction
of a string field theory.
We had expected that associated with the physical zero-momentum dilaton
state there was a
CFT deformation. While the
non-primary nature of the dilaton creates complications, it is possible to
write a deformation associated with the dilaton. The
deformation, however, turns out to be trivial. The CFT
is not deformed at all.  More concretely, the deformed correlators can
be made to agree with the original correlators by a redefinition of
the local operators of the theory. The redefinition is generated by
the ghost number operator, and precisely reproduces the deformation
induced by the dilaton by virtue of the ghost number anomaly.
This means that the zero-momentum dilaton is not really a parameter
of the conformal theory.\foot{The effect is  analogous to the
one where the $\theta$ parameter of QCD becomes unobservable in the
presence of massless fermions.} Nevertheless the dilaton turns out
to deform a string background.  While we do not give an axiomatic
definition of a string background we sketch a possible definition
in sect.~7.2. To show that two apparently different
string backgrounds are identical we must show that that all the
forms can be made to coincide by a redefinition of the local operators
of the conformal theory. Since forms on moduli spaces involve antighost
operators, the similarity transformation induced by
the ghost number operator, that
cancels out the deformation for the conformal theory, does not cancel
out the deformation of the string background. Thus the dilaton provides
an example where the {\it space of vacua of string theory has a parameter
that is absent in the space of conformal field theories.} This implies
a subtle difficulty in attempting to use
two-dimensional theory space
as the precise arena for the formulation of a manifestly
background-independent string theory. Further investigation of the
difference between a CFT and a string background
is likely to be fruitful. A string background that arises
from a CFT will be called a conformal string background.
It is certainly conceivable that there are string backgrounds
that do not arise from CFT's.

At any rate, since string backgrounds related by a dilaton deformation
are different,
we can ask whether two string field theories constructed using
two such backgrounds are the same. Proving they are the same amounts to
showing that these backgrounds are
backgrounds of the {\it same}
string field theory.  In this light it is clear that the dilaton theorem
is a case study for background independence.
It is a novel case because
the different backgrounds originate from the same conformal theory.

In the analysis of background independence in Ref.[\senzwiebachtwo], the
fields to be shifted were BRST physical, primary fields of the form
$c\bar c\O$, with $\O$ a primary,
dimension $(1,1)$ field constructed out of the matter sector of the CFT.
Since the dilaton is not primary, in order to
insert it in some region of a Riemann surface we need
a family of local coordinates, namely a local
coordinate for every point in the region.
Such families of local coordinates can be obtained by introducing metrics
on the surfaces. To extract local coordinates from metrics we use a
prescription given by Polchinski [\polchinski]. Such local coordinates
are known in the mathematical
literature as the $\omega$-normal coordinates associated to a K\"ahler metric
with K\"ahler form $\omega$ (see Ref.[\gilkey], sect.~3.7). Here we will simply
call them local normal-coordinates. Choosing the metrics
that must be used to get the local normal-coordinates is a delicate matter.
For this we must extend some of the results of Ref.[\senzwiebachtwo] concerning
the $\K$ operator and the $\B$ spaces.
The $\K$ operator acting on a surface $\Sigma$, equipped with
coordinate disks $D_i$, gives the two dimensional space of surfaces
represented by the surface $\Sigma$ with a new puncture
inserted anywhere on $\Sigma-\cup D_i\,$. Acting on a space of surfaces,
$\K$ will do this to each surface in the space. The local coordinate
at the new puncture is not specified. $\B$-spaces,
the new geometrical ingredient appearing in background independence, arise
roughly as homotopies induced by $\K$ ($\B_{g,n}$ provides
a homotopy between $\K\V_{g,n-1}$ and $\V_{g,n}$). Since $\B$ spaces
always have a special puncture, and the local coordinate at this puncture
is not defined, the recursion relations satisfied by the $\B$ spaces
are weak equalities, namely, they hold up to the local coordinate at the
special puncture.
We define an operator $\ov\K$ that
adds a puncture and specifies the local coordinate
at the puncture. Acting on a surface, the operator $\ov\K$
uses the minimal area metric on that surface to extract the local
normal-coordinates.
We then show that the recursion relations for the
$\B$ spaces can be turned into strong equalities, needed for
dealing with dilatons. While in Ref.[\senzwiebachtwo] $\B$-spaces
had to have at least two punctures, one special and one ordinary, we find
it useful to introduce higher genus $\B$ spaces with one
special puncture and no ordinary puncture.

We use the results of Ref.[\belopolskyzwiebach] to compute the two-form
associated with a $\d$ surface-insertion that uses the most
general family of local
coordinates. Similarly, we compute the one-form associated with a $\x$
line-insertion. Particular cases of this result were obtained in
Refs.[\polchinski,\distlernelson,\nelson].
When the family of local coordinates
is the family of local normal-coordinates associated with
a metric on the surface, the
dilaton two-form reduces to the curvature two-form of the metric,
in agreement with
a result obtained earlier in Ref.[\polchinski]. Under the same circumstances
we show that a line integral of the $\x$ one-form over a curve
precisely computes the integral of the geodesic curvature over the curve.
Since both the curvature two-form, and the geodesic curvature are elements of
the Gauss-Bonnet theorem expressing the Euler number of a surface in
terms of the Riemannian data on the surface, we see that both the
$\d$ two-form and the $\x$ one-form must be present in a regularized version
of \distnel. For a surface $\Sigma$ of genus
$g$ with $n$ punctures, and equipped with coordinate disks $D_i$, the
regulated form of Eqn.~\distnel\
 takes the form, schematically,
$$\Bigl\langle\, \Bigl( \hskip-6pt\int_{\Sigma-\cup D_i}\hskip-14pt d^2z
 D(z,\bar z) + \hskip-18pt\int_{\p(\Sigma-\cup D_i)}\hskip-18pt d\xi\,
 \chi (\xi) \,\Bigr) \Phi_1(P_1)\cdot\cdot \Phi_n(P_n)
\Bigr\rangle_{\Sigma}\sim\,
(2\hskip-2pt -\hskip-2pt 2g\hskip-2pt -\hskip-2pt n)
\,\langle \Phi_1(P_1)\cdot\cdot \Phi_n(P_n)
\rangle_{{\Sigma}} \,. \eqn\rdistnel$$
Here the dilaton is only integrated over the surface
minus the unit disks defining the coordinates at the punctures, and
the $\chi$ operator is integrated over the boundaries of
the unit disks. The expression is perfectly regulated since operators
never collide.
Unlike \distnel,
the states inserted at the punctures
need not be physical; this is essential in order to go off-shell.
The insertions of $D$ and $\chi$ are done with the local normal-coordinates
arising from an arbitrary metric on $\Sigma$.\foot{It should be possible
to show that \rdistnel\ holds for a general family of local coordinates,
including those not arising from  metrics. We will not attempt to do this
in the present work.}
We use the $D$ and $\chi$ forms to construct the CFT deformation related
to \rdistnel, and to develop
its generalization to the cases when
we integrate over
the positions of the external states and over the moduli of the surface.

The paper is organized as follows. Section 2 begins with a review of the
basic properties of the dilaton state and of the procedure for
extracting families of local normal-coordinates from metrics. We then review
the BV (Batalin-Vilkovisky)
algebra of Riemann surfaces and the
corresponding BV
algebra of string functionals.
In section 3 we discuss modified versions of the $\K$ operator
that insert
a puncture on Riemann surfaces and define a local coordinate for the
insertion. We also introduce an operator $\L$ that,
acting on a surface, inserts a puncture on the coordinate curves.
We review the properties of the homotopy
spaces $\B$, and show how to obtain strict recursion relations taking into
account the local coordinates at the inserted punctures.
In section 4 we first motivate and then give
a precise formulation of the dilaton theorem, indicating what needs to
be proven. In section 5 we begin by reviewing the
Gauss-Bonnet theorem in two dimensions, paying particular attention to
the contribution of geodesic curvature. We then compute the
dilaton two-form and the
$\x$ one-form for the case of a general family
of local coordinates. When the local coordinates arise from a metric, the
dilaton two-form and $\x$ one-form are shown to become the curvature
two-form and the  geodesic
curvature respectively.
In section 6 we derive two general results concerning the
insertion of the dilaton and of the state $\x$ in spaces of Riemann surfaces.
The insertions are performed using the operators $\K$ and $\L$.
For $\x$ we derive another result, describing its insertion
in a space of surfaces by twist-sewing a three-punctured sphere carrying
$\x$ onto each surface of the space.
In section 7 we show explicitly how to construct a CFT
deformation using the dilaton and $\x$. We then prove that the deformation
is actually trivial. We sketch a definition of a string background, and
argue that the dilaton deforms it nontrivially.
In section 8 we prove the dilaton
theorem by constructing an antibracket preserving diffeomorphism of
the state space $\H$ into itself,  mapping into each other the
string field measures $d\mu e^{2S/\hbar}$ built
with two different but nearby values of the dimensionless string coupling
constant. The inhomogeneous piece of the diffeomorphism is a shift
along the dilaton direction. The nonlinear pieces of the diffeomorphism
use both the $\B$ spaces and the state $\x$.

\chapter{Background Material and Preliminary Developments}

This section has three parts. In the first part we review
the definition of the ghost-dilaton state $\d$,
introduce the $\x$ state, and give their main properties. This
material is
taken mostly from Ref.[\distlernelson],\foot{For other relevant
work on dilatons see Refs.[\nelson,\imbimbo,\wong] .} and is summarized
here for convenience and in order to set up our conventions.
In the second part we review the concept of a family of local
coordinates. We show explicitly how to extract a family of local
coordinates from a conformal metric following Ref.[\polchinski].
This prescription is familiar in the mathematical literature and
applies to any K\"ahler metric [\gilkey]. Finally, in the third part,
comprising sections 2.3--2.5, we review the
BV algebra of Riemann surfaces,
the construction of forms on moduli spaces, and the map from spaces
of surfaces to string functionals.

\section{Basics of the Dilaton}

In bosonic string theory formulated
on a background spacetime
which is twenty six dimensional Minkowski space, the dilaton
vertex operator reads
$$D_p(z,\bar z) = \Bigl( \,c\bar c \,\partial X \cdot\overline\partial X
- {1\over 2}\, ( c \partial^2 c - \bar c \overline \partial^2 \bar c ) \Bigr)
\, e^{ipX}\, .\eqn\vopdil$$
This state is well-known to be BRST
invariant if the on-shell condition
$p^2=0$ holds. Therefore the dilaton represents a physical massless scalar.
Note that the ghost structure of the operator
is rather nontrivial. Despite being physical, this operator is not primary.
Indeed the operator product expansion with the holomorphic part
of the total stress tensor is given by
$$T(z) D_p(w,\bar w) \sim {1\over (z-w)^3} \Bigl( -ic\bar c p\cdot \bar
\partial
X  + c\partial c \Bigr) \, e^{ipX}\, + \cdots \; .\eqn\opedil$$
While this only establishes that the above representative of the
dilaton BRST cohomology class is not primary, it is also known that there is no
primary
representative valid for all values of the momentum satisfying $\,p^2=0$.
Primary representatives for the dilaton
have difficulties with the value $p=0$.
The above representative for the dilaton is no exception and
is not primary when $p=0$. In this case the violation of the primary
state condition arises from the ghosts.
It is therefore of interest to focus on the ghost part of the dilaton
vertex operator. The zero-momentum ghost-dilaton is present for any
consistent string background, and is defined by
$$D(z,\bar z) \equiv
\, {1\over 2} \, ( c \partial^2 c - \bar c \overline \partial^2 \bar c )\,
\,  ,\eqn\vpdil$$
or in the Fock space version, by
$$\d =  (c_1c_{-1} - \bar c_1 \bar c_{-1} ) \ket{0}\,. \eqn\dilfock$$
This state is still annihilated by the total BRST operator $Q$.
An instructive way to verify this goes as follows.
One first checks that
$$D(z,\bar z) =\, {1\over 2}
 \{\, Q , \partial c -\bar \partial \bar c \,\}\,,  \eqn\qghost$$
and therefore one can write
$$\d = \, {1\over 2}\,\{ Q , \partial c -\bar \partial \bar c \} \ket{0}=
Q c_0^- \ket{0}= - Q\, \x \,\, , \eqn\abstriviall$$
where we have introduced the state $\x$ defined as
$$\x \equiv\,   -\,c_0^-\ket{0}
=\,- {1\over 2}\, (c_0 - \bar c_0) \ket{0}\,.
\eqn\abstrivial$$
The representation of $\d$ in \abstriviall\ makes it manifest that $Q \d =0$.
In addition, the dilaton is annihilated by $b_0^-(\equiv b_0-\bar b_0)$
$$b_0^-\d = b_0^- Q c_0^- \ket{0} = L_0^- c_0^-\ket{0} - Q \{ b_0^-, c_0^-\}
\ket{0} = 0\,,\eqn\versubs$$
and is therefore an element of the restricted state space defined by
$$\H = \{\ket\Psi\in\HH : b_0^-,L_0^-\ket\Psi = 0\}\; ,\eqn\restricted$$
making it a perfectly acceptable physical state of
the closed string spectrum.
The zero-momentum ghost dilaton is a trivial state in the absolute
BRST cohomology, but is not trivial in the
semirelative BRST cohomology that properly
defines the closed string physical states. This is so because
$\d$ is given by $Q$ acting on the
state $\x$, which is  not annihilated by $b_0^-$,
$$b_0^-\x =\, -\,\ket{0}  \not= 0\,, \eqn\nogood$$
and therefore, is not part of the restricted state space $\H$.

As previously stated, the dilaton is not primary, but as expected from
\opedil\ only $L_1$ and
$\ov L_1$ fail to annihilate it
$$ \{ L_0, L_2, L_3, \ldots   \} \d = 0 , \quad
 \{ \ov L_0, \ov L_2,\ov  L_3, \ldots   \} \d = 0,\eqn\system$$
while
$$  L_1 \d \not= 0 \, ,\quad\hbox{and}\quad \ov L_1 \d \not= 0\,
.\eqn\violpri$$
Indeed,
$$\eqalign{
L_1\ket{D} =  \{ Q , b_1\} \ket{D} &= Q b_1
(c_1c_{-1} - \bar c_1 \bar c_{-1} ) \ket{0}\cr
&= -Qc_1\ket{0}= -\{ Q, c\}\ket{0}
= -c \partial c \ket{0} = c_0c_1\ket{0}\,,\cr} $$
and similarly $\ov L_1 \d = -\bar c_0 \bar c_1 \ket{0}$. The $\x$ state
is also not primary, it fails to be so just as much as the
dilaton
$$ \{ L_0, L_2, L_3, \ldots   \} \x = 0 , \quad
 \{ \ov L_0, \ov L_2,\ov  L_3, \ldots   \} \x = 0\, , \eqn\xsyst$$
while
$$  L_1 \x \not= 0 \, ,\quad\hbox{and}\quad \ov L_1 \x \not= 0\, .\eqn\viopri$$
Indeed, one easily verifies that
$$L_1\x =  \{ Q , b_1\} \x =  b_1 Q \x = -b_1\d =  c_1 \ket{0}\,,\eqn\examp$$
and, similarly, $\ov L_1 \x = -\bar c_1 \ket{0}$.

The representative we have chosen for the zero-momentum dilaton
state is not unique. We could have written, for example,
$\d_\alpha = -Q c_0^-\ket{0} + \alpha \, Q c_0^+\ket{0}$,
with $\alpha$ an arbitrary constant. The second term here represents a
truly BRST trivial term. We will find no use in the present work for
these representatives and we will use the representative
$\d$ throughout.

\section{Family of Local Coordinates from a Conformal Metric}

A family of local coordinates defined
over a region $R$ of a Riemann surface is a
prescription that assigns a local coordinate around every point in the
region $R$.
In other words, for each point $p\in R$ we know how to
fix a canonical disk around it.
We assume we have a uniformizer $z$ for the
region $R$, and the region is itself parameterized by two real parameters
$\lambda_1$ and $\lambda_2$. This means that a point $p\in R$ defines
uniquely the values of $\lambda_1$ and $\lambda_2$, and there is a function
$z(\lambda_1,\lambda_2) = z(p)$. Having a family of local coordinates
amounts to having a family of maps
$h_{\lambda_1,\lambda_2} : D \to R$, taking a unit disk $D$
into the region $R$,
such that the center of $D$ is mapped to the point $p\in R$ corresponding
to
the parameters $\lambda_1$ and $\lambda_2$ (see Figure 1).
We can describe explicitly the family of local
coordinates as
$$z = h_{\lambda_1\lambda_2} (w) = z(\lambda_1,\lambda_2) +
a(\lambda_1,\lambda_2)\, w +  \, {1\over 2}\, b(\lambda_1,\lambda_2)\,w^2 +
{1\over 3!} c(\lambda_1,\lambda_2)\, w^3  + \cdots \; ,\eqn\famcoor$$
where $w\in D$.
The origin of the disk is mapped to the point $p$
specified by the parameters $\lambda_1, \lambda_2$ because
$ h_{\lambda_1\lambda_2} (0) = z(\lambda_1,\lambda_2) = z(p)$.
Equation \famcoor\ describes the most general family of local coordinates
in the region $R$. This family is specified by the functions
$z(\lambda_1,\lambda_2),\, a(\lambda_1,\lambda_2),\, b(\lambda_1,\lambda_2)$,
and so on.

\if y\figcount
\midinsert
\epsfxsize 4.5in
\centerline{\epsffile{figure1.eps}}
\nobreak
\narrower
\singlespace
\noindent
Figure 1. The point $p\in R$ is parameterized by $\lambda_1,\lambda_2$.
The map $h_{\lambda_1,\lambda_2}$ defines a local coordinate around the
point $p$, by mapping the unit disk to some neighborhood of $p$ such
that the origin of the disk is mapped to $p$.
\medskip
\endinsert
\else\fi

Given a metric $\rho(z,\bar z)$ in a domain $R$ described by the local
uniformizer $z$, one can extract
in some canonical fashion a family of local coordinates for the domain $R$.
One method was proposed in Ref.[\senzwiebachtwo]: the coordinate disk
around each point $p$ is defined to be the locus of all points whose shortest
distance to $p$ is less than or equal to some fixed number $a_0$. While
this prescription is fairly intuitive it does not lend itself so easily
to explicit computation. We will use a related prescription given in
Ref.[\polchinski]. Fix a point $p\in R$ and let the
function $z = h_{\lambda_1\lambda_2} (w)$, with
$z(p)= h_{\lambda_1\lambda_2} (w(p))= h_{\lambda_1\lambda_2} (0)$,
define the local coordinate $w$ to be used for the point $p$.
In this prescription $h_{\lambda_1\lambda_2} (w)$ is determined by
pulling  back the metric to the $w$ plane, and demanding that the pulled
metric $\rho^w$ satisfy $\partial_w^n \rho^w |_{w=0} =
\partial_{\bar w}^n \rho^w|_{w=0} =0$,
for all positive values of $n$. This prescription, of course, does not
determine the parameterization $z(\lambda_1,\lambda_2)$ which must be
supplied. It also does not determine  the first
coefficient $a(\lambda_1,\lambda_2)$ of the series expansion. The absolute
value $|a|$ can be fixed easily, but the phase of $a$ cannot be fixed in
any straightforward way. All higher order coefficients are
uniquely determined.

We begin our computation by relating the metric $\rho^z$ referred to
the uniformizer $z$ to the metric $\rho^w$ referred to the $w$ coordinate,
$$\rho^z |dz| = \rho^w |dw| , \quad \to \quad
\rho^w = \rho^z\Bigl|{dh_\lambda\over dw}\Bigr|\, . \eqn\getm$$
Our aim is to use the above equation to express the metric in the $w$ plane
as a power series in $w$ and $\bar w$. Expanding the right hand side we find
$$\rho^w = \Bigl[ \rho^z|_p + (z-z(p)) \,\partial_z\rho^z|_p + (\bar z
-\overline {z(p)}\,)\, \partial_{\bar z}
\rho^z|_p + \cdots \Bigr]\cdot
 \Bigl|a+ bw + \cdots\Bigr|\, ,\eqn\pullmetric$$
where we only keep terms linear in $w$.
The coordinate difference $(z-z(p))$ can be expressed in terms of $w$
using the definition of the family of local coordinates \famcoor ,
$$\rho^w = \Bigl[ \rho^z|_p + aw \,\partial_z\rho^z|_p +
\bar a \bar w\, \partial_{\bar z}
\rho^z|_p + \cdots \Bigr]\,  |a|\,  \Bigl( 1 + {b\over 2a} w +
{\bar b\over 2\bar a} \bar w+\cdots \Bigr) \,.\eqn\sympli$$
Collecting all terms of order $w$ (or $\bar w$) we find
$$\rho^w =  |a|\, \rho^z|_p \,\,
\Bigl[  1+  aw \,\Bigl( \partial_z\ln\rho^z|_p+ {b\over 2a^2} \Bigr)
 + \bar a \bar w\, \Bigl( \partial_{\bar z}
\ln\rho^z|_p + {\bar b\over 2\bar a^2} \, \Bigr) \,+ \cdots \Bigr]\,.\eqn\gty$$
The prescription now requires that
$\partial_w \rho^w|_{w=0} = \partial_{\bar w} \rho^w|_{w=0} =0$, giving us
$${b\over 2 a^2} =  -\partial_z\ln\rho^z|_p\, , \quad\hbox{and}\quad
{\bar b\over 2 \bar a^2} =  -\partial_{\bar z}\ln\rho^z|_p\, .\eqn\almfr$$
Since these equations (consistent with complex conjugation, since $\rho$
is real) determine the ratio $b/a^2$ for each point $p\in R$ we can
simply write
$$ {b\over 2a^2} =  -\partial_z\ln\rho^z \, .\eqn\finaldet$$
This equation will be quite useful later on. As mentioned earlier, $|a|$
can be fixed naturally by setting $|a| = 1/\rho^z|_p$. In this way the
metric $\rho^w$ becomes of the form
$\rho^w = 1 + \cdots$, as can be seen from \gty.
The procedure does not fix the phase of $a$ over the region $R$.
We will learn how
to fix that phase
in a natural way for the case of families of local coordinates
over curves.

\section{Algebra of Riemann Surfaces}

We review in this section some of the geometrical results of Ref.
[\senzwiebachtwo].
The following spaces of Riemann surfaces will be of use to us :
$\M_{g,n}$ denotes the usual moduli space of Riemann surfaces of genus $g$ and
$n$ punctures, $\ov\M_{g,n}$ denotes its compactification (which includes
the degenerate surfaces), $\P_{g,n}$ denotes the space of Riemann surfaces
of genus $g$ and $n$ punctures with local coordinates defined at each
puncture, and $\wh\P_{g,n}$ is the same as $\P_{g,n}$ except that the phases
of the local coordinates are not specified. The spaces $\P_{g,n}$ and
$\wh\P_{g,n}$ are fiber bundles over $\ov\M_{g,n}$, and $\P_{g,n}$ is a
circle bundle over $\wh\P_{g,n}$.
We will be dealing mostly with
$\wh\P_{g,n}$, since this is the space
that is directly relevant in string field theory.

\noindent
\underbar{Algebra of Riemann Surfaces}
Local coordinates allow us to twist-sew two punctures by identifying points
whose
local coordinates are related by
$$z_1z_2=t\; ,$$
where $t= \exp (i\theta)$, with $\theta \in [0, 2\pi ]$.
Given two {\em symmetric}\foot{Under the exchange of the labels at the
punctures.} subspaces $\A_1\subset\wh\P_{g_1,n_1}$ and
$\A_2\subset\wh\P_{g_2,n_2}$, the subspace
$\{ \A_1 ,  \A_2\} \subset\wh\P_{g_1+g_2,n_1+n_2-2}$ is defined
by twist sewing every surface in $\A_1$ to every surface in $\A_2$,
and symmetrizing over the $(n_1 + n_2 - 2)$ remaining
punctures. We can use any two fixed (labeled) punctures
to sew since the subspaces are symmetric.
This operation possesses the following properties
$$\{\A_1 , \A_2\}= -\,(-)^{(\A_1+1)(\A_2+1)} \,\{\A_2,\A_1\}\,, \eqn\exchange$$
$$ (-)^{(A_1+1)(A_3+1)} \Bigl\{ \{ \A_1 , \A_2\}\,  , \, \A_3\Bigr\}\,+\,
\hbox{cyclic} \, = 0\,,\eqn\jcbdntty$$
$$ \p\, \{\A_1 ,  \A_2\}  = \{ \p\A_1 ,  \A_2\} + (-)^{\A_1 +1} \{ \A_1 ,
\p\A_2 \} \; , \eqn\ebountimes$$
where $\A_1$ and $\A_2$ in the exponents denote the dimensions of the spaces.
Given a symmetric subspace $\A\subset\wh\P_{g,n}$, the
subspace $\Delta\A \subset\wh\P_{g+1,n-2}$ is defined
by twist sewing two fixed (labeled) punctures on
every surface in $\A$, and multiplying by a factor of a half.
Together with the antibracket defined above this operation satisfies
$$\eqalign{
\Delta^2 \A &= 0\,, \cr
\Delta  \{ \A_1, \A_2\}&=\,\,\{\Delta\A_1,
\A_2\}+(-)^{\A_1+1} \{\A_1,\Delta\A_2\}\,,\cr
\Delta\p\A &= -\p\Delta\A\; .\cr }\eqn\edeltasquare$$
If one defines a complex $\wh\P$ as the formal sum, over $g$ and $n$,
of the spaces $\wh\P_{g,n}$,
the antibracket and delta operator are well defined in this
complex.

Occasionally we will underline a symmetric subspace $\underline\A$ to denote
that one of its punctures has been selected as a special puncture.
Since the space is symmetric, the choice of puncture is irrelevant.
This puncture is reserved for the insertion of some definite external
state, and is therefore never sewn.
The subspaces $\{\underline\A,\B\}$
and $\Delta\underline\A$ are then defined as above except that the special
puncture is ignored altogether.
For the antibracket, symmetrization
is performed only over the
{\em ordinary} punctures,
with the special puncture fixed on $\underline\A$. Consequently
$\{\underline\A,\B\}$ is only symmetric under the exchange of ordinary
punctures.

\section{Forms on Moduli Spaces}

The operator formalism of CFT assigns a state
$\bra\Sigma\in (\HH^*)^{\otimes n}$ to every surface
$\Sigma\in\P_{g,n}$.
This allows one to define canonical forms in
$T_{\Sigma}(\P_{g,n})$ by the rule\foot{The notation
differs slightly from
reference [\senzwiebachtwo] in the way the degree of the form is
specified. In  [\senzwiebachtwo] $\Omega^{(k)g,n}$ denoted a
$(6g-6+2n+k)$-form,
here $\Omega^{[k]g,n}$ will denote a $k$-form.
We changed the rounded parentheses
to brackets to eliminate possible confusion, if both notations were to
be used simultaneously.}
$$\bra{{\Omega_{}^{}}^{[k]g,n}}( V_1,\cdots , V_k )
\equiv (-2\pi i)^{(3-n-3g)}\bra{\Sigma}\,{\bf b}({\bf v}_1)\cdots
{\bf b}({\bf v}_k).
\eqn\cdefform$$
The Schiffer vector ${\bf v}_r= (v_r^{(1)}(z),\cdots v_r^{(n)}(z))$
creates the deformation of the surface $\Sigma$ specified by the tangent
$V_r$, and the antighost insertions are given by
($\ointop dz/z=\ointop d\bar z/\bar z=2\pi i$)
$${\bf b}({\bf v}) = \sum_{i=1}^n \biggl(
\oint b^{(i)}(z_i) v^{(i)}(z_i) {dz_i\over 2\pi i}
+\oint \overline b^{(i)}( \overline z_i)  \overline v^{(i)}
(\overline z_i) {d\overline z_i\over 2\pi i} \biggr).\eqn\kjhkjh$$
These forms satisfy the familiar $Q\leftrightarrow d$ correspondence,
$$ \bra{\Omega^{[k]g,n}} \sum_{i=1}^n Q^{(i)}
=(-1)^k d\bra{\Omega^{[k-1]g,n}} .\eqn\eketbrst$$
When the states inserted at the $n$ punctures are in $\H$ these
forms descend to well-defined forms on $T_\Sigma \wh\P_{g,n}$
[\nelson,\zwiebachlong].

We now develop an application of the above results that will be useful in our
later developments. We will calculate the explicit antighost insertions
associated with {\it tangents that represent the motion of a puncture in a
domain
$R$ of some surface}. As the puncture moves the local coordinate at that
puncture changes. We will also assume that as this motion takes place
the rest of the surface data (moduli, position of the other punctures, and
their local coordinates) does not change.
Therefore the Schiffer vector
will be supported only on the moving puncture. The associated antighost
insertion will be necessary to study the effect of insertions
of $\d$ and $\ket\chi$ in correlators.

The local coordinate at the moving puncture is described
by the  family of local coordinates \famcoor\
over a region $R$ of a Riemann surface (with uniformizer $z$)  .
Associated with the real parameter $\lambda_i$ there is a Schiffer vector
$v(\lambda_i)$ given by [\belopolskyzwiebach]
$$v_{\lambda_i}(w) = -{dh_{\lambda_1\lambda_2} (w)\over d \lambda_i}
\Bigl[{dh_{\lambda_1\lambda_2}(w)\over dw}\Bigr]^{-1}\; .\eqn\schif$$
Evaluation gives
$$v_{\lambda_i}(w) = \alpha_{\lambda_i} + \beta_{\lambda_i}w +
\gamma_{\lambda_i}w^2
+\cdots\,\, ,  \eqn\schifv$$
where the first three coefficients are given by
$$\eqalign{
\alpha_{\lambda_i} &= -{1\over a} {dz\over d\lambda_i}\,\,, \cr
\beta_{\lambda_i}&= -{1\over a}  {da\over d\lambda_i} +
{b\over a^2} {dz\over d\lambda_i}\,\,, \cr
\gamma_{\lambda_i} &= -a  \,{d\over d\lambda_i}\Bigl(
{b\over 2a^2} \Bigr)
- \Bigl( {b^2\over a^3}
- {1\over 2} {c\over a^2} \Bigr) {dz\over d\lambda_i}\; .   \cr}\eqn\coeff$$
The antighost insertion corresponding to this Schiffer vector is then
$$\eqalign
{ {\bf b}(v_{\lambda_i}) &= \,\,\,\, \alpha_{\lambda_i}b_{-1}
\,+\,\beta_{\lambda_i}b_0
\,+\,\gamma_{\lambda_i}b_1 + \cdots  \cr
&\quad + \overline\alpha_{\lambda_i}\,\bar b_{-1}
+\overline\beta_{\lambda_i} \,\bar b_0
+\overline\gamma_{\lambda_i}\, \bar b_1 + \cdots\; , \cr}\eqn\binsert$$
where the bars denote complex conjugation on the coefficients,
and  antiholomorphic sector on the antighosts.

\section{Maps to String Functionals}

Integration of the the canonical forms over subspaces of $\wh\P_{g,n}$
defines a map from these subspaces to
functions on the vector space $\H$. One takes
$$f(\A_{g,n}^{[k]}) \,\equiv {1\over n!}\,
\int_{\A_{g,n}^{[k]}}\bra{\Omega^{[k]g,n}}\Psi\rangle_1\cdots
\ket{\Psi}_n\,\, , \quad n\geq 1\, ,\eqn\fgeomty$$
where $\A_{g,n}^{[k]}$ is a $k$-dimensional subspace,
and the subscripts on the string field insertions denote which copy
of $\H$ they belong to.
In the space $\H$ there is a natural symplectic structure.
The antibracket and delta operator are given by [\senzwiebachtwo]
$$\eqalign{
\bigl\{ f \, , g \, \bigr\} &=\, (-)^{g+1}
{\p \,f \over \p \,\ket{\Psi}}\, {\p \,g\over \p \,\ket{\Psi}}\,\ket{\s}\, ,\cr
\Delta \, f &= (-)^{f+1} \Bigl(\, {\p\over\p\ket{\Psi}}
{\p\over\p\ket{\Psi}}\, f \Bigr) \, \ket{\s}\, ,\cr}\eqn\takedelta$$
where $(-)^g$ denotes the grassmanality of $g$,
and the twist-sewing ket $\ket{\s}\in\H^{\otimes 2}$ is
given by
$$\ket{\s_{12}} = b_0^{-(1)}\ket{R_{12}'} =  b_0^{-(1)}\int {d\theta\over 2\pi}
  e^{i\theta L_0^{-(1)}}\ket{R_{12}} \; ,\eqn\sewingket$$
where the reflector ket $\ket{R_{12}}$ is defined by
$\bra{R_{12}}R_{23}\rangle ={}_3{\bf 1}_1 \;$. The twist-sewing ket
satisfies $\bra{\omega_{12}}\s_{23}\rangle = {}_3{\bf 1}_1$, where
$\bra{\omega_{12}}\equiv \bra{R_{12}'} c_0^{-(2)}$ is the symplectic
form in $\H$.

It then follows that for $\A, \B \,\subset\wh\P$ the following
representation identities hold [\senzwiebachtwo],
$$\eqalign{
f\bigl( \Delta \A ) &= -\Delta  f(\A) \, \cr
f \bigl( \{ \A , \B \} \bigr) &= - \{\, f(\A) , f(\B) \}\, .\cr}\eqn\hreh$$
The above equations define a homomorphism from the
BV algebra on subspaces
of $\wh\P$ to a  BV
algebra on string field functionals in $\H$, a
fact that is formulated in a precise fashion in Ref.[\senzwiebachnew]. From
\eketbrst\ it follows that
$$\{ \,S_{0,2} , f (\A )\} = \, - \,f\bigl( \p \A\bigr)
\, ,\eqn\reppd$$
where $S_{0,2}$ is the kinetic term of the string field action, given by
$$S_{0,2} = {1\over 2}\,\bra{\Psi} c_0^- Q \ket{\Psi} = {1\over 2}
\,\bra{\omega_{12}}
Q^{(2)}\ket{\Psi}_1\ket{\Psi}_2\,. \eqn\kinterm$$

\noindent
\underbar{Maps with Insertions.}
More general maps can be defined with the use of specific
states in the state space of the conformal theory. We will denote
by $\ket{\O}$  states in $\HH$ and by $\ket{\wh\O}$ states in the
restricted space $\H$. The state may have arbitrary grassmanality and
arbitrary ghost number.
We will only consider
functionals with one additional insertion, defined as
$$f_\O(\A_{g,n+1}^{[k]}) \,\equiv
{1\over n!} \int_{\A_{g,n+1}^{[k]}} \bra{\Omega^{[k]g,n+1}\,}
\Psi\rangle_1\cdots\ket{\Psi}_{n}\ket{\O}_{n+1}\, ,\eqn\fstate$$
and similarly for states $\ket{\wh\O}$.
If $\A$ is a symmetric space the state can be inserted anywhere,
but if there is one puncture that is not symmetrized over the
state must be inserted at that puncture.
For general subspaces of $\wh\P$ we must only use
$\ket{\wh\O}$ states.
This condition is well known to make the insertions well-defined
regardless of the phase of the local coordinate at the insertion point.
This is important because typically there are no sections
in $\P_{g,n}$ over general
subspaces of $\wh\P_{g,n}$, namely, one cannot fix the phases of the local
coordinates smoothly over general subspaces of $\wh\P_{g,n}$.
However, in subspaces over which the phase of the
local coordinate at the special puncture can be defined
continuously, we can insert $\ket{\O}$ states  without ambiguities.
In the following section we will introduce a large class of such subspaces,
the simplest of which are spaces of three-punctured spheres.
These spaces will
later be used to insert the unrestricted state $\x$.

The following identities can be shown to hold
$$\eqalign{
f_\O \bigl( \Delta \A ) &= -\Delta  f_\O(\A) \,, \cr
f_\O \bigl( \{ \A , \underline{\B} \} \bigr) &=
- \{\, f(\A) , f_\O(\B) \}\, ,\cr
\{ \,S_{0,2} \,,\, f_\O (\A)\} &= \, - \, f_\O(\p\A)\, + \,
  (-)^\A  f_{Q\O}(\A) \,  ,\cr}\eqn\hrehn$$
where the underline denotes the space in which the extra state is inserted.
The first two identities were given in [\senzwiebachtwo], for the case
of grassmann even states in $\H$. Note that the special puncture in
the surfaces contained in the $\A$ and $\B$ spaces must
have a well defined phase.
For each of the above equations there are trivial cases. If the
surfaces in  $\A$ have
less than three punctures the first equation gives zero equal zero.
If  the surfaces in $\A$
have no punctures, or those in $\B$ have less than two punctures,
the second equation is
trivial. Finally, if the surfaces in $\A$ have only one puncture
the left hand side of the third
equation vanishes identically, and the right hand side vanishes by cancelation,
$$  f_\O(\p\A_{g,1})\, = \,
    (-)^{\A_{g,1}}f_{Q\O}(\A_{g,1}) \, .\eqn\importantr$$
It is useful to introduce, for states in $\H$, the odd Hamiltonian function,
$${\bf U}_{\wh\O (0,2)}\equiv \bra{\omega_{12}\,}\wh\O\rangle_1\ket{\Psi}_2
\,\, . \eqn\shftpsi$$
Note that for states $\ket{\O}$
annihilated by $c_0^-$, this definition would give zero.
This function implements the insertion of the state $\ket{\wh\O}$ by a
canonical transformation,
$$f_{\wh\O} (\A) = \,\{\, f(\A)\, , \,{\bf U}_{\wh\O
(0,2)}\,\}\,\,.\eqn\createp$$
Furthermore,
$$ \{ \, S_{0,2} \,,\, {\bf U}_{\wh\O(0,2)} \} = -{\bf U}_{Q\wh\O(0,2)}
\,,\eqn\vnshcon$$
which vanishes if the state $\ket{\wh\O}$ is physical.
Given that the string action $S$ can be written as
$$S= S_{0,2} + f(\V) + \hbar S_{1,0}\,, \eqn\saction$$
where $S_{1,0}$ is string field independent and $\V$ are the string
vertices (to be reviewed briefly in sec.~3.2), we find
$$\{ S, f_\O (\A) \} = \,(-)^\A f_{Q\O} (\A)  -
f_\O \Bigl(  \,\p\A +
\{ \V , \underline\A \} \Bigr) \,,  \eqn\excellent$$
where use was made of the last two equations in \hrehn. Notice that if
the surfaces in $\A$ have just one puncture the above equation is still
valid, and becomes zero equal zero by virtue of \importantr.
Again, the special puncture of the surfaces in $\A$ must have a
coordinate with a
well-defined phase whenever the state $\ket{\O}$ is outside $\H$.
Another useful relation
follows from \createp\ and \vnshcon\ ,
$$ \{ \, S \,,\, {\bf U}_{\wh\O(0,2)} \} = -{\bf U}_{Q\wh\O(0,2)} +
f_{\wh\O} (\underline\V)
\,\,.  \eqn\beexcellent$$

\chapter{Adding Punctures and $\B$-spaces}

The present section is an extension of some of the work in
Ref.[\senzwiebachtwo].
As far as Riemann surfaces are concerned, the main tools introduced in
[\senzwiebachtwo] were
an operator $\K$ that inserts a new puncture into surfaces, and a series
of spaces $\B$ satisfying geometrical recursion relations. Since
Ref.[\senzwiebachtwo] was concerned with deformations by primary fields,
there were simplifications that are not possible when we consider the
dilaton. We will therefore refine the discussion by introducing several
classes of $\K$ operators, a new insertion operator $\L$, and $\B$ spaces with
one puncture.

\section{Adding a puncture}

An operator $\K$ that adds punctures to Riemann surfaces
was defined in [\senzwiebachtwo].\foot{In [\senzwiebachtwo] this operator was
actually denoted as $\wh\K$.}
Given a surface $\Sigma \in \wh\P_{g,n}$,
we define a two-dimensional subspace $\K (\Sigma)\subset\wh\P_{g,n+1}$
which contains the $(n+1)$-punctured surfaces given by $\Sigma$ with an
extra puncture lying anywhere in the region $\Sigma-\cup_i D_i$,
where $D_i$ are unit disks around the original punctures. The local
coordinate at the extra puncture is fixed arbitrarily but continuously
over the relevant region.
For a subspace $\A\subset\wh\P_{g,n}$, we define the subspace
$\K\A\subset\wh\P_{g,n+1}$ as the collection of spaces $\K (\Sigma)$
for all $\Sigma\in\A$.
We will refer to such subspaces as $\K$-spaces.
If the subspace $\A$ is symmetric then $\K\A$ will
be symmetric in its first $n$ punctures. The inserted puncture is
special in that it is not symmetrized over and is never sewn.
The arbitrariness in choice of the local coordinate at the special puncture
necessitates the introduction of the concept of {\em weak equality}. Two
subspaces
$\A,\B\subset\wh\P_{g,n}$, containing each one special puncture,
are said to
be {\em weakly equal}, $\A\approx\B$, if they are equal up to the local
coordinates at the special punctures. One can show that the following
weak identities hold [\senzwiebachtwo],
$$\K\, (\,\{ \A_1 , \A_2\}\, )\, \approx\, \{ \K\A_1 , \A_2\} \,+\,
\{ \A_1 , \, \K \A_2 \} \eqn\com$$
$$\K \, (\Delta \, \A) \approx \Delta\, (\K \A)\eqn\bvdelta$$
$$[\,\partial\,,\,\K\,]\,\approx\,-\,\{\, \V'_{0,3}\, , \,\, \}\,,\eqn\xz$$
where $\V'_{0,3}$ is the 3-punctured sphere with one special puncture
introduced
in [\senzwiebach]. It is symmetric under the exchange of two of its punctures.
The third, special puncture, is the one associated with $\K$.
The above weak equalities were sufficient for the discussion of background
independence,
since the states inserted at the special puncture were always
primary. For the dilaton we need strong equalities. This is so
because the dilaton is not primary and
its insertion depends on the local coordinate at the special puncture.
\foot{When we insert states such as $\x$, we must even fix the phase of
the local coordinate at the special puncture.}

We can fix the local coordinate at the special puncture
using the metric which solves the minimal area
problem of closed string field theory.\foot{The conformal metric of
least area under the condition that all nontrivial closed curves be longer than
or equal to $2\pi$.} Given a surface $\Sigma$ equipped with coordinate
disks $D_i$ around the punctures, we find the minimal area metric on $\Sigma$
and use the prescription of sec.~2.2 to extract local coordinates at every
point on $\Sigma -\cup D_i$. These are the coordinates to be used for inserting
the extra puncture, and the corresponding
insertion operator is denoted as $\overline\K$.
Recall that the minimal area metric gives a canonical way of choosing
coordinate disks on any punctured surface.
\foot{For such a metric some neighborhood of each puncture
is isometric to a semi-infinite cylinder of circumference $2\pi$. The
coordinate
curve is chosen to be the geodesic on the cylinder a distance $\pi$
from the boundary of the cylinder.} If the punctured surface $\Sigma$
comes equipped with its own coordinate disks, these may or may not coincide
with the coordinate disks that the minimal area metric on $\Sigma$
would determine.
It follows that the operation $\ov\K$ of
adding an extra puncture using the minimal
area metric, is particularly natural acting on surfaces
whose coordinate curves are chosen using the minimal area metric.
For such spaces of surfaces, denoted as $\V_i$,
we find the strong identities
$$\eqalign{
\overline\K\, (\,\{ \V_1 , \V_2\}\, )\, &=\, \{ \overline\K\V_1 , \V_2\}
 \,+\,\{ \V_1 , \, \overline\K \V_2 \} \,,\cr
\overline\K \, (\Delta \, \V_i) &= \Delta\, (\overline\K \V_i)\, \,,\cr
[\,\partial\,,\,\ov\K\,]\,&=\,-\,\{\, \V'_{0,3}\, , \,\, \}\,.\cr}
\eqn\comdeltap$$
The first two equations hold by virtue of the compatibility of minimal
area metrics with the operation of sewing.
The last equation holds acting on $\V_i$ spaces,
when the coordinate at the special puncture
of $\V'_{0,3}$ is induced by the metric representing the
underlying {\em symmetric} two punctured sphere as a flat infinite cylinder of
circumference $2\pi$.

As minimal area metrics have curvature singularities, extracting local
normal-coordinates can be delicate. One can
smooth out the singularities by changing the metric on the surface
minus its unit disks, keeping it unaltered in some neighborhood
of the unit disks. The deformed metric will still be
flat and isometric to that of a
cylinder of circumference $2\pi$ throughout the unit disks.
This metric is no longer unique, as there
are many ways of deforming the minimal area metric on the surface.
The associated insertion operator will be denoted $\K^*$.  In a sense,
this is the operator most appropriate for the dilaton.
Note that for these metrics sewing is still isometric gluing since
the coordinate curves are still geodesics of the metric.
Moreover, the bulk curvature
of these metrics on the surface minus its disks is not changed from its
original value.
As we will see at the end of sect.~6.1,
while the first two equations in \comdeltap\
only hold weakly for $\K^*$, they will actually hold strongly when a
dilaton is inserted at the special puncture. The third equation in
\comdeltap\ holds strongly for $\K^*$ since it involves only
the coordinate curves of the
surfaces, where the metric is unaltered.

\noindent
\underbar{The $\L$ operator} It is convenient to introduce an operator
$\L$ that, acting on a surface $\Sigma \in \wh\P_{g,n}$,
gives a one-dimensional subspace $\L (\Sigma)\subset\wh\P_{g,n+1}$
which contains the $(n+1)$-punctured surfaces given by $\Sigma$ with an
extra puncture lying anywhere on the
coordinate curves $\p(\Sigma-\cup D_i)$.
The local coordinate at the extra puncture is fixed arbitrarily but
continuously
on the curves.
For a subspace $\A\subset\wh\P_{g,n}$, we define the subspace
$\L\A\subset\wh\P_{g,n+1}$ as the collection of spaces $\L (\Sigma)$
for all $\Sigma\in\A$.
We will refer to such subspaces as $\L$-spaces.
If the subspace $\A$ is symmetric then $\L\A$ will
be symmetric in its first $n$ punctures.
The orientation of $\L\A$ is
defined as follows. Let $\{ \A \}$ denote the orientation of $\A$
at some point corresponding to the surface $\Sigma$, and let the
special puncture be at some point on one of the coordinate curves.
Let $v$ denote
the tangent vector to the coordinate curve at that point, with the
coordinate curve oriented as the boundary
of $\Sigma -D$, and let
$\wh V(v)$ be the corresponding tangent in $\wh\P_{g,n+1}$
representing the motion of the special puncture. The orientation of
$\L\A$ is then defined as $[\wh V(v), \{ \A \} ]$. For more details the reader
may consult sect.~2.3 of Ref.[\senzwiebachtwo].

The definition of $\L$ helps us clarify Eqn.~\xz. Just by virtue of our
definitions, the commutator of the boundary operator $\p$ and $\K$
is precisely $\L$,
$$[\,\p \, , \, \K\, ] = \L\, \,, \eqn\relkl$$
as one readily verifies by
acting on a space of surfaces. It then follows
from \xz\ that $\L \approx -\{\V_{0,3}'\, , \, \}$.
If we use a
particular metric
on the surfaces of a space $\A$ to define the local coordinates for the
insertion $\K$ we will use the same metric for the insertion $\L$.
It then follows that $\ov\L$, defined using the minimal area metric, satisfies
$$[\,\p \, , \, \ov\K\, ] = \ov\L\,, \,\,\,\quad  \ov\L =
 -\{\V_{0,3}'\, , \, \}\,, \eqn\relko$$
where the second equation holds acting on $\V_i$ spaces by virtue of the
last equation in \comdeltap.

If the metric used by $\L$ to extract the local
coordinate at the special puncture is invariant under rigid $U(1)$ rotations
of the coordinate curve, we expect that the $\L$ insertion can be
reproduced by twist-sewing a
three-punctured sphere $\V^*_{0,3}$ whose
local coordinate at the special puncture is chosen to
match the one used by $\L$.
If the metric used by $\L$ is not $U(1)$ invariant there is no
guarantee that the $\L$ space can be obtained by twist sewing.

$\L$ spaces are useful since one can define the phase of the
local coordinate at the special puncture {\em uniquely} for each
surface $\Sigma\in\L\A$, and {\em continuously}
over the space $\L\A$.
This does not quite define a section in $\P$, since we are only fixing
the phase at the special puncture, but it does allow the insertion of an
unrestricted state (like $\x$) at the special puncture. The phase is
defined by identifying it with the direction of the tangent to the
coordinate curve where the special puncture is inserted.
Continuity  over $\L\A$ is
is guaranteed by uniqueness, and the continuity of the coordinate curves
over $\A$.

\section{The Homotopy Spaces $\B$}

The string vertices of string field theory $\V_{g,n}$ are sections in
$\wh\P_{g,n}$ over compact subsets
of the moduli spaces $\M_{g,n}$ [\zwiebachlong].
They are all assembled together as
$$\V\equiv\sum_{g,n}\hbar^g\kappa^{2g-2+n}\,\V_{g,n}\; ,
\quad\hbox{with}\quad
\cases{n\geq 3 \,\,\hbox{for}\,\,g=0 \cr
n\geq 1  \,\,\hbox{for}\,\,
g= 1\,\cr
n\geq 0 \,\,\hbox{for}\,\,
g\geq 2\; .} \eqn\strvert$$
Note that vacuum vertices $\V_{g,0}$ exist for $g\geq 2$, but not for genus
one. We have also included in the sum above the relevant powers
of the string field coupling
constant $\k$ (set equal to one in [\senzwiebachtwo]).
The string vertices satisfy the following recursion relations
$$\partial \V + \hbar \Delta \V +
{1\over 2}\{ \V ,\V\} =0\, \quad .\eqn\sumrecursions$$
A comment is in order here. While the vacuum vertices are not relevant
to the master equation, they are relevant to the action.
The recursion relations
\sumrecursions, originally written only for vertices
with punctures [\sonodazwiebach] , were
later recognized [\zwiebachlong] to determine the vacuum vertices,
since the operations
$\Delta \V$ and $\{ \V , \V \}$ produce surfaces without punctures even
if all the surfaces in $\V$ had punctures. Thus \sumrecursions\ gives
constraints for vacuum vertices. Since the operations
$\Delta \V$ and $\{ \V , \V \}$ cannot produce surfaces of genus
one without punctures (recall that at genus zero the lowest vertex has
three punctures) it is not natural to include a vacuum vertex
$\V_{1,0}$.\foot{In Ref.[\senzwiebach] a genus one vacuum vertex
was defined while discussing the background independence of the
genus one contribution to the free energy. Still, there was no canonical
choice for this vacuum vertex.}

A set of subspaces $\B_{g,n}\subset\wh\P_{g,n}\,$ of dimension
$(6g+2n-6)+1$ were introduced in
[\senzwiebachtwo]:
$$\B\equiv\sum_{g,n}\hbar^g\kappa^{2g-2+n}\,\B_{g,n}\,\quad\hbox{with}\quad
\cases{n\geq 3 \,\,\hbox{for}\,\,g=0,\cr
n\geq  2 \,\,\hbox{for}\,\,
g\geq 1\,.}\, \eqn\bspaces$$
Each space $\B_{g,n}$ must have one special puncture, and is
symmetric in all others. The marginal operator, or the dilaton in
this paper, is inserted at the special puncture. The string field
is to be inserted at the other punctures.
The requirement that $\B$ spaces have at least one puncture other than the
special one was imposed since
these spaces were
used to construct Hamiltonian functions, and
field independent terms in the Hamiltonian drop out from
background independence conditions.
The subspaces are defined recursively as homotopies between the symmetric
string vertices $\V$ and new spaces $\V'$ (see Eqn.~(5.1) of [\senzwiebachtwo])
$$\p\B \simeq \V' - \underline\V \,,\quad\hbox{with}\quad \V' = \k\V'_{0,3} +
\k\K\V- \hbar\Delta\B - \{\V,\B\}  \,. \eqn\rwrt$$
The symbol $\simeq$ means that the equality holds up to the local coordinates
at the special puncture, {\em and} up to surfaces of genus one with one special
puncture and no ordinary puncture. Note that we have inserted the required
factors of $\k$.

We need  better control over the $\B$ spaces since we will insert dilatons
on them, and dilatons capture some information about the local coordinates
at the special puncture. We first claim that apart from surfaces having
no ordinary punctures \rwrt\ can be made to hold strongly by careful
definition of the way we insert the extra puncture. Thus we
claim that in the sector where $n\geq 2$ (and $n\geq 3$ for $g=0$)
we can find $\B$ spaces such that
$$\bigl( \p\B\, \bigr)_{n\geq 2}= \Bigl( \k\V'_{0,3} +
\k\ov\K\V- \hbar\Delta\B - \{\V,\B\} - \underline\V\Bigr)_{n\geq 2}
\,\,. \eqn\improveb$$
Note that we have replaced the operator $\K$ which used no specific
way of introducing coordinates at the extra puncture by $\ov \K$
which
introduces coordinates using the minimal area metric on the surface.
This equation can be established recursively. Note that we can proceed
order by order in the dimensionality of the $\B$ spaces since the boundary
of an $r$-dimensional $\B$ space on the left-hand side of the equation
can only involve
$\B$ spaces of dimensionality lower than $r$ on the right-hand side. The unique
$\B$ space of lowest dimension is the one-dimensional space $\B_{0,3}$.
To lowest order this equation requires
$\p\B_{0,3} = \V'_{0,3}- \underline\V_{0,3}$.
Since both $\V'_{0,3}$ and $\V_{0,3}$ have well defined
coordinates around the special puncture, we can clearly find an
interpolating $\B_{0,3}$ with well-defined local coordinates around
the special puncture. To guarantee we can solve \improveb\ recursively
we must establish that acting on the right-hand side of this equation with $\p$
must give zero, since the right hand side is supposed to be the boundary
of a space. Since $\V'_{0,3}$ is just a point, we must show that
$$\Bigl( \k\,\p\,\ov \K\V +\hbar\Delta\p\B-\{\p\V,\B\} + \{
\V,\p\B\}-\p\underline{\V}\,\Bigr)_{n\geq 2} \, = 0\, . \eqn\sptro$$
In order to find out if this equation holds we need to use the
expression for $\p\B$. We can actually use the expression in
\improveb\ since the terms in $\p\B$ with no ordinary punctures, of which we do
not keep track, cannot contribute to the left hand side of \sptro. This
left hand side then becomes
$$\eqalign{\Bigl[  & \,-\k\,\{ \ov\K\V,\V\} - \hbar\k\,\Delta\ov\K\V - \,\k\,
\{ \V'_{0,3},\V\} \cr
&\quad +\hbar (\,\k\,\Delta \V'_{0,3} + \k\Delta\ov\K\V - \{\Delta\V,\B\}
+\{ \V,\Delta\B\} -
\Delta\underline{\V}) \cr
&\quad +{1\over 2}  \{ \{ \V,\V\},\B\} + \hbar\{ \Delta \V ,\B\} \cr
&\quad+ \{ \V\, ,\, \k\,\V'_{0,3} + \k\ov\K\V - \hbar\Delta\B -
\{\V,\B\}-\underline{\V}\}\cr
&\quad+ \{ \underline{\V} ,\V \} + \hbar
\Delta\underline{\V}\,\Bigr]_{n\geq 2}\; , \cr} \eqn\themess$$
where we made use of the strong equalities listed in \comdeltap, and
for the help of the reader each term in \sptro\ has been written as
one line in \themess. Note that the term $\hbar\k\Delta \V'_{0,3}$
appearing in the second line,
does not carry an ordinary puncture, hence it
must be dropped.
Using the
exchange property of the antibracket \exchange,
and the Jacobi identity \jcbdntty,
it is easily checked that all terms in \themess\ cancel out. This
verifies the consistency condition. Thus the $\B$ spaces, with $n\geq 2$
can be defined recursively as homotopies between two spaces with coincident
boundaries (see Ref. [\senzwiebach] for the construction of
homotopies). This proves \improveb.

Since the $\B$ spaces do not contain
surfaces with just one puncture,
Eqn.~\rwrt\ implies that
$$(\V' -\underline\V )_{g,1} \approx 0 \, \quad \hbox{for} \quad  g\geq 2
\,.\eqn\recalls$$
Since all the $\B$ spaces were defined by the
recursive construction above, Eqn.~\recalls\ must be verified to hold.
This was done in Ref.[\senzwiebachtwo], sec.~5.2. We do not think
that equation \recalls\ can be made to hold strongly by any choice
of the $\B$ spaces. Thus we will use the failure of \recalls\ to hold strongly
to define $\B$ spaces with $n=1$, that is $\B$ spaces
with no ordinary punctures.
We attempt to set
$$\p \B_{g,1} \equiv (\V' -\underline\V )_{g,1} \,
=\ov\K\V_{g,0}- \sum_{g_1=1}^{g-1}\{ \V_{g_1,1}\, , \,
\B_{g-g_1, 2} \}
- \Delta \B_{g-1, 3} - \underline\V_{g,1}\,,
\quad  (g\geq 2)\; .\eqn\calls$$
The consistency of this definition requires that
$$ \p (\V' -\underline\V )_{g,1}  = 0\, ,\eqn\neednow$$
and this equation can be verified to hold strongly with a calculation
similar to the one we did in \themess.\foot{Since the verification just
involves $\B$ spaces with $n\geq 2$, whose identities now hold strongly, the
computation of [\senzwiebach], Eqn.~(5.9) suffices.} It now follows that
with our extended definition
$$\B\equiv\sum_{g,n}\hbar^g\kappa^{2g-2+n}\,\B_{g,n}\,\quad\hbox{with}\quad
\cases{n\geq 3 \,\,\hbox{for}\,\,g=0 \cr
n\geq  2 \,\,\hbox{for}\,\,g= 1\,\cr
n\geq  1 \,\,\hbox{for}\,\,g\geq 2}
\; , \eqn\bspaces$$
we have that
$$\p\B\, = \k\V'_{0,3} +
\k\ov\K\V- \hbar\Delta\B - \{\V,\B\} - \underline\V
\,\,, \quad  (g,n) \not= (1,1) \; .\eqn\improv$$
Note that the consistency of this equation is guaranteed because
the $n=1$ $\B$-spaces just introduced
do not contribute to the right hand side ($\Delta\B_{g,1} = 0$
and $\{\B_{g,1},\A\}= 0$).
The only piece of information that is not encoded
in the above strong equation is given in \recalls\ and now takes the form
$$\p\B_{g,1} \approx 0\,  , \quad (g\geq 2) \,. \eqn\extra$$

We could try to eliminate the restriction to $(g,n) \not= (1,1)$ in
\improv.
The right hand side of \rwrt\ contains surfaces of genus one without an
ordinary puncture.
It is therefore
tempting to introduce the space $\B_{1,1}$ such that
$\p \B_{1,1} \equiv(\V' -\underline\V )_{1,1}=
-\Delta\B_{0,3} - \underline\V_{1,1}$.
This equation is not consistent since
$\p^2\B_{1,1} \not=0$. This is not surprising, since in the
analysis of [\senzwiebachtwo] the conditions of
background independence of the free energy at genus one involved theory
space connections, in addition to Riemann surfaces.

We can still fix \improv\ by hand, by subtracting from the right
hand side the offending terms. We then have
$$\p\B\, = \k\V'_{0,3} +
\k\ov\K\V- \hbar\Delta(\B- \k \B_{0,3}) - \{\V,\B\} - \underline\V
+ \hbar\k\,\underline\V_{1,1}\, \eqn\improvet$$
which is a strong equality without any qualifications. It is now useful
to define
$$\eqalign{ \B_> &\equiv \B - \k \B_{0,3}  \cr
\V_> &\equiv \V - \k \V_{0,3}\,, }\eqn\introdnew$$
and then rewrite the recursion relations as
$$\p\B_>\, =
\k\ov\K\V- \hbar\Delta\B_> - \{\V,\B\} - \underline\V_>
+ \hbar\k\,\underline\V_{1,1}\, .\eqn\improvett$$
This is also a strong equality without qualifications.

\chapter{The Dilaton Theorem}

In this section we begin by giving some motivation for the dilaton theorem.
We explain what goes wrong when one tries to implement a dilaton shift
as a gauge transformation generated by the state $\x$. This
argument gives some insight into the kind of string field redefinition
necessary to prove the dilaton theorem. We then turn to the precise statement
of the dilaton theorem in string field theory, and explain what must be
proven.

\noindent
\underbar{Motivation}
Consider an infinitesimal shift of the string field by the dilaton state $\d$.
Since $\d = -Q \x$ this can be written as
$$\ket{\Psi} \to \ket{\Psi} -  {\epsilon\over \k}\, Q\x \, .\eqn\fakegt$$
This shift has the form of an infinitesimal gauge transformation,
except that it is
not a legal one, given that $b_0^- \x \not= 0$.
Nevertheless it leaves the kinetic term of the action invariant,
since
$$\k \cdot \delta S_{0,2} \sim \bra{\Psi}  c_0^- Q Q\x +
\bra{D} c_0^- Q \ket{\Psi}= \bra{D}c_0^-Qb_0^-c_0^-
\ket{\Psi}=0 \; .$$
The second equality follows by inserting the identity $1=\{b_0^-,c_0^-\}$, and
the
last equality follows by anticommuting $b_0^-$ with $Q$ and using
$(Q, b_0^-, L_0^-)\ket{D} = 0$.
Since \fakegt\ has succeeded to be a gauge transformation for the
kinetic term of the closed string field theory, it is of interest to see
if we can write a full nonlinear gauge transformation of the classical theory.
The candidate gauge parameter is clearly $-{1\over \k}\x$, and the naive
gauge transformation would read [\zwiebachlong]
$$\eqalign{
\delta_\chi \ket{\Psi} &=  {\epsilon\over \k}\, \d  +
{\epsilon\over \k}\sum_{n=3}^\infty {\k^{n-2}\over (n-2)!} \bra{V^{(n)}}
\Psi\rangle^{n-2}\cdot
\x \ket{\s} \cr
&=\epsilon\,\Bigl\{\,
\ket{\Psi}\,,\, {1\over\kappa}\,{\bf U}_{D(0,2)} + \sum_{n=3}^{\infty}
\kappa^{n-3}f_{\chi}(\underline\V_{0,n})\Bigr\} \; .\cr}
\eqn\fakegtnl$$
In the second step we use the results presented in section 2.5 to
write the gauge transformation as a canonical transformation.
The first nonlinear term corresponds to an insertion of the unrestricted state
$\x$ on a three string vertex.
Since $L_0^- \x =0$, and no antighost insertion is necessary, this term is
independent of the choice of phase that is made at the special puncture.
We have already seen that the leading variation of the action, of order
$\O(\k^{-1})$, vanishes. In addition, the $\O (\k^0)$ terms in the variation of
the action also cancel,
$$\{S_{0,2},f_\chi (\underline\V_{0,3})\} +
\{f(\V_{0,3}),{\bf U}_{D(0,2)}\} = 0 \; .\eqn\sthvar$$
The second equality follows from equations \hrehn\ and \createp, since
$\p\V_{0,3} = 0$.

Since $\x$ is not in $\H$, in order to define the higher
terms unambiguously, we must choose
continuously over $\V_{0,n}$ a phase for the coordinate around the
puncture where this state is inserted. In other words we must find a section
in $\P_{0,n}$ over $\V_{0,n}\subset\wh\P_{0,n}$.
While this is not obviously
feasible, it actually turns out that it can be done.\foot{One thinks of the
punctures as moving in the plane, with three of them fixed at $0,1$ and
$\infty$. One can then take the phase of each puncture
to be measured from the positive $x$-axis, for example. There is no global
problem since the punctures never collide in $\V_{0,n}$. One cannot fix,
however, the phase around a puncture in the higher genus $\V_{g,n}$ vertices,
due to topological complications.
There are no global sections in these cases.}
Despite the existence of sections in $\P_{0,n}$
over $\V_{0,n}$, the $n\geq 4$ terms in \fakegtnl\
{\em do not lead to an invariance of the action}.
The origin of the problem is the failure of the {\em sections} over $\V_{0,n}$
to satisfy the geometric recursion relations \sumrecursions, which
are crucial for proving gauge invariance of the classical action.

\if y\figcount
\midinsert
\epsfysize 2.5in
\centerline{\epsffile{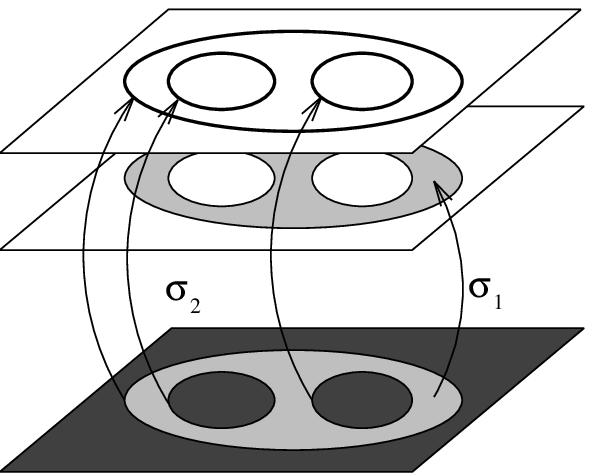}}
\nobreak
\vskip -1.75in \hskip 0.5in ${\cal P}_{0,4}$
\nobreak
\vskip .8in \hskip 0.5in $\overline{{\cal M}}_{0,4}$
\nobreak
\vskip .25in
\narrower
\singlespace
\noindent
\vskip 5pt
Figure 2. The lowest plane represents $\overline{\M}_{0,4}$, or
equivalently its section in $\wh\P_{0,4}$. The dark shaded regions
represent the Feynman diagrams formed by sewing two three-vertices
$\V_{0,3}$, and the light shaded region is $\V_{0,4}$.
The maps $\sigma_1$
and $\sigma_2$ give the sections in $\P_{0,4}$ over $\V_{0,4}$ and over
the boundaries of the Feynman regions respectively.
The section $\sigma_1$ cannot be deformed such that its boundary
matches the section $\sigma_2$.
\medskip
\endinsert
\else\fi

Let us consider the example of four-punctured spheres. Their compactified
moduli space $\ov\M_{0,4}$ is itself a sphere. Three of the punctures are
fixed, and the fourth is thought of as moving on this sphere. A local
coordinate with a well-defined phase at the position of the fourth puncture
defines a nonvanishing vector with a unique direction at that position; the
direction  corresponding to zero phase.
As the puncture moves on the sphere it defines a vector field, and the
Poincar\'e-Hopf theorem requires such a vector field
to have an index equal to two.
In string field theory
the space $\ov\M_{0,4}$
is decomposed into four regions: the
region $\V_{0,4}$, and three disjoint regions corresponding to the
Feynman diagrams
formed by sewing two three-vertices $\V_{0,3}$.
The earlier choice of phases in  $\V_{0,3}$ actually fixes the vector
field in each of the Feynman regions. It turns out that in
each of the Feynman regions the vector field has an index equal to $+1$.
Let us just assume that
the index in each Feynman region is some integer $i_F$  (by symmetry it
must be the same for the three regions). Since one must fix the phase
continuously in $\V_{0,4}$, the  vector field in this region
will have index zero. Consequently, if the four vector fields defined
over the four regions could be patched continuously to obtain
a single vector field over the sphere, this resulting vector field
would have index $3i_F$.
Since the vector field must have index two, this is impossible.
Therefore the phase at the boundaries $-{1\over 2} \{ \V_{0,3},\V_{0,3}\}$
of the Feynman regions, cannot be made to agree with the phase at
$\partial\V_{0,4}$, and the geometric recursion relation fails to hold in
$\P_{0,4}$ (see Figure 2).

Since a dilaton shift cannot be extended to a true gauge transformation,
we must look for a string field transformation that brings the action
back to a recognizable form. That diffeomorphism of the string field
will be seen to change the coupling constant of the theory. Since the
first two terms of the pseudo-gauge transformation had the correct effect
of preserving the action to order $\O (\k^0)$,
we suspect those two terms will be present in the general
diffeomorphism
$$\delta\,\ket{\Psi} = \epsilon\,\Bigl\{ \,\ket{\Psi}\,,
{1\over\kappa}\,{\bf U}_{D(0,2)}
+ f_{\chi}(\underline\V_{0,3}) + \cdots \Bigr\}\, .\eqn\truncgt$$

\noindent
\underbar{Precise Formulation}
In analogy with the proof of background independence [\senzwiebachtwo],
the dilaton theorem can be formulated precisely as an off-shell statement
in string field theory.
Consider two identical copies $\H_\k$ and $\H_{\k'}$ of the state space
$\H$. These copies are labeled by the value of the dimensionless
coupling constant to be used in the construction of the
corresponding string field
theory. There should be a symplectic diffeomorphism
$$F: \H_{\k'} \to \H_\k\,\,, \eqn\sympldiff$$
that pulls back the relevant
{\em measure} of a theory with coupling constant $\kappa$ to
the measure of a theory with coupling constant $\kappa '$, namely,
$$ F^* \left\{ d\mu (\k) \exp \left({ 2\over \hbar} S (\k) \right)
\right\} \, =d\mu(\k ') \exp \left({ 2\over \hbar} S (\k ') \right)
\, .\eqn\nud$$
This means that an antibracket preserving field redefinition, which
we expect to include a constant inhomogeneous term, maps a string theory with
one coupling constant to a string theory with another coupling constant.
It follows that the two string theories are just two different backgrounds
of the same theory.

For the case when the two coupling constants differ only infinitesimally,
$$\k' = (1+ a\e ) \k \, \eqn\shfcc$$
with $a$ some constant to be fixed, an infinitesimal symplectic diffeomorphism
must implement the desired map between the theories,
$$F :  \ket{\Psi} \,\to\, \ket{\Psi}
 + \epsilon\{\ket{\Psi} ,{\bf U}_D\} \, , \eqn\infi$$
where ${\bf U}_D$ is an odd Hamiltonian function. Our intuition about
the dilaton tells us that  ${\bf U}_D$ must begin by implementing a shift
along the dilaton
$${\bf U}_D = {1\over \k} {\bf U}_{D(0,2)} + \cdots\, ,
\quad\to \quad F : \ket{\Psi} \to \ket{\Psi} + {\e\over \k}\d + \cdots
\,.  \eqn\guesst$$
We must find the complete form of ${\bf U}_D$ and the value of the
constant $a$. This will establish the infinitesimal form of the dilaton
theorem.

It is interesting to consider the classical limit, where
\nud\ requires $S (\k')  = F^* S(\k)$, with $S$ now the classical
master action (the limit of master action as $\hbar$ goes to zero).
This  means that
$$S\, \Bigl(\,\k+ a\e\k\, , \ket{\Psi}\, \Bigl)  =
S\, \Bigl( \, \k\,\,, \ket{\Psi} + {\e\over \k} \d + \cdots \Bigr)\,.
\eqn\classlimit$$
This equation says that at the classical level a change of the string
coupling constant can be achieved by giving the string field some
expectation value in the direction of the dilaton (and performing a
field redefinition).

Let us now derive the explicit form of the equation that ${\bf U}_D$
must satisfy.
If we write
$d\mu (\k) = \rho(\k) \prod d\psi$, one can show
that [\senzwiebachtwo]
$$F^* (d\mu(\k)) = {\rho(\k)\over \rho(\k ')}\, d\mu(\k ')
\left( 1 + 2\epsilon\Delta{\bf U}_D\right) \, .\eqn\measure$$
Moreover,
$$F^* \{ S(\k)\} = S(\k) + \epsilon\, \{ S(\k) , {\bf U}_D \} \,.\eqn\tractd$$
Equation \nud\ then reduces to
$$a \k \, {d\over d\k} \left( S + {1\over 2} \hbar \ln \rho\right)
 =  \{ S , {\bf U}_D \} +
\hbar \Delta {\bf U}_D \, .\eqn\mainc$$
This is the equation that must be satisfied by
the odd Hamiltonian ${\bf U}_D$, for some value of the constant $a$.
This equation is rather similar to the equation expressing the condition
of background independence of the string field theory (Eqn.~(4.16) of
Ref.[\senzwiebachtwo])
$$\hbox{D}_\mu(\widehat\Gamma) \left( S + {1\over 2} \hbar \ln \rho\right)
-{1\over 2} \hbar \hbox{str}\, \widehat\Gamma_\mu =
\{ S , {\bf U}_\mu \} + \hbar \Delta {\bf U}_\mu \; ,\eqn\maincbi$$
where the aim was to find the Hamiltonian ${\bf U}_\mu$.
Since the dilaton state is not marginal, the dilaton
Hamiltonian ${\bf U}_D$ is expected to be somewhat
more complicated than the
background independence Hamiltonian, which could be expressed fully in
terms of $\B$ spaces and the marginal operator as
${\bf U}_\mu= {\bf U}_{\mu\,(0,2)} - f_\mu (\B)$.
As discussed before Eqn.~\truncgt\ we expect that
$${\bf U}_D = {1\over\k}{\bf U}_{D(0,2)} + f_{\chi}(\underline\V_{0,3})
+ \cdots \, ,\eqn\missing$$
and finding the missing terms will be our objective. The full diffeomorphism
will involve $\B$ and $\V$ spaces, and both $\d$ and $\x$.
This will be derived in sect.~8. The correct value of $a$ turns out
to be $a=+1$.

Using the form of the string action in \saction\ and the expression
\strvert, the condition \mainc\ for the dilaton theorem can be written
more explicitly as
$$\eqalign{
\{ S , {\bf U}_D \} +
\hbar \Delta {\bf U}_D &= a\hbar \k \, {d\over d\k}
\left(  S_{1,0} + {1\over 2}  \ln \rho\right) \cr
&\quad + a \sum_{g,n}(2g\hskip-1pt -\hskip-1pt 2+n)\,\hbar^g\k^{2g-2+n}
\,f(\V_{g,n}) \, ,\cr} \eqn\maine$$
where the sum over $g$ and $n$ runs precisely over the nonvanishing
string vertices (Eqn.~\strvert). We have used the fact that the
kinetic term $S_{0,2}$ is independent of $\k$. It is not clear to us
whether the genus one terms included in the first line of the right
hand side should be there. The naive expectation is that they should
have no coupling constant dependence, but this expectation might not
be realized.

\chapter{Topology and Dilaton two-forms}

In this section we will begin by reviewing the Gauss-Bonnet theorem
for the case of two dimensional Riemannian
manifolds with boundaries. This theorem gives the Euler number
in terms of an integral of curvature over the
manifold and an integral of geodesic curvature over the boundaries.
We will then calculate the general expressions for the dilaton two-form,
and the $\x$ one-form when a generic family of local coordinates is used
to produce the insertions.
We then consider the case of a family
of local normal-coordinates arising from a conformal metric on the
surface. We show that the dilaton two-form becomes the curvature two
form, a result familiar from Ref.[\polchinski], and the $\x$ one-form
is seen to compute geodesic curvature. The latter result requires the
definition of the phase of the local coordinate at the
boundary of the surface. This can always be done in a canonical fashion.

\section{Gauss-Bonnet Theorem}

For a two dimensional Riemannian manifold $M$ with a smooth
boundary $\p M$, the
Gauss-Bonnet theorem states that the Euler number $\chi(M)$ of the manifold
can be written as [\singer]
$$ \chi (M) = {1\over 2\pi} \int_M R^{(2)} \, + \, {1\over 2\pi}
\int_{\partial M} k \,  ,
\eqn\gauss$$
where $R^{(2)}$ is the curvature $2$-form, and $k$ is geodesic curvature.
For a genus $g$ surface $\Sigma$ with $n$ boundary components
the Euler number is $\chi (\Sigma) = 2-2g-n$.
Writing the area element as $dA = \rho^2 dx\wedge dy =
{i\over 2}\rho^2 dz\wedge d\bar z$,
the curvature $2$-form is given by
$$\eqalign{
R^{(2)} &= - {4\over\rho^2}\,\partial\bar\partial\ln\rho \, dA \cr
&=  - 2i\,\partial\bar\partial\ln \rho\, (dz\wedge d\bar z) \cr
&= id \,\Bigl( d z \,  \partial \ln \rho \,
-\,  d\bar z \, \bar\partial \ln \rho \Bigr)\, .\cr}
 \eqn\curvature$$
It is of interest to give an explicit expression for the geodesic
curvature $k$. The basic property of this curvature is that given an oriented
open curve $\gamma$ with origin $x_i$ and endpoint $x_f$, embedded in a two
dimensional manifold with a metric, the
integral $\int_\gamma k$ of the geodesic curvature measures the
rotation angle
from the tangent $T(x_i)$ parallel transported to $x_f$ to the tangent
$T(x_f)$ at $x_f$. When the curve is a geodesic we get zero,
since geodesics are defined by parallel transporting
the (unit) tangent vector.
It is possible to find a local formula for the geodesic curvature
$k$ in a region $R$ described by the uniformizer $z$. We find
$$k= \hbox{d} \theta_\gamma  - i\,[\, d z \,  \partial \ln \rho \,
-\,  d\bar z \, \bar\partial \ln \rho] \, .\eqn\geodcurv$$
In this expression the term  $i[\cdots ]$ computes the rotation angle of
any vector after parallel transport from $z$ to $z + dz$. The first term
$\hbox{d} \theta_\gamma$ computes the rotation angle of the tangent to
the curve $\gamma$. The difference computes the angle from the
parallel-transported tangent to the new tangent.
Note that $k$ depends on the choice of curve $\gamma$. Consequently
$k$ is a well-defined one-form on this curve, but not on the manifold
$M$ in which it sits. At each point $p\in M$ one finds many geodesic
curvature one-forms, depending on the chosen curve through that point.
Acting with the exterior derivative on $k$ produces a two-form that is
well-defined on $M$, and is given precisely by the curvature
$$ dk = -R^{(2)} \; .\eqn\rtrivial$$

When a two dimensional manifold $\Sigma \in \V_{g,n}$
is equipped with a minimal area
metric, the integral of bulk curvature over the surface
minus its unit disks $(\Sigma-\cup D_i)$  is given by
$${1\over 2\pi}\int_{\Sigma-\cup D_i} R^{(2)} =
\chi(\Sigma-\cup D_i)  = 2-2g-n\, .\eqn\euler$$
This follows because the coordinate curves
of a punctured surface equipped with a minimal area metric are geodesics of
that metric (see sect.~3.1).
Therefore, there is no geodesic curvature contribution to
\gauss.

\section{Dilaton two-form and $\x$ one-form}

Consider once again the family of local coordinates \famcoor\
over a region $R$ of a Riemann surface,
defined by
$$z =  h_{\lambda_1\lambda_2} (w) = z(\lambda_1,\lambda_2) +
a(\lambda_1,\lambda_2)\, w +   {1\over 2} b(\lambda_1,\lambda_2)\,w^2 +
{1\over 3!} c(\lambda_1,\lambda_2)\, w^3  + \cdots\, \,,\eqn\famcoor$$
where $\lambda_1$ and $\lambda_2$ denote two real parameters parameterizing
the domain $R$, which has a uniformizer $z$.
We now define a dilaton two-form representing the measure
for integrating a dilaton insertion
$$ \ket{\omega_{D}^{(2)}} \equiv d\lambda_1\wedge d\lambda_2 \,
{\bf b}(v_{\lambda_1})
\,{\bf b}(v_{\lambda_2})\,\d\, . \eqn\dilmea$$
Note that this is a two-form on the domain $R$, living in the state space
of the conformal theory. The motivation for this definition is that
$\ket{\omega_{D}^{(2)}}$ is clearly an ingredient in the
construction of the canonical forms given in \cdefform. It is now a
short computation using \dilfock, \coeff, and \binsert, to show that
the two-form $\ket{\omega_{D}^{(2)}}$ only has a component along the
vacuum state of the CFT,
$$ \ket{\omega_{D}^{(2)}} = \ket{0}\cdot\omega_D^{(2)} \, , \eqn\dilform$$
where the two-form $\omega_D^{(2)} $ on the region $R$ is given by
$$\eqalign{
\omega_{D}^{(2)} &= d\lambda_1\wedge d\lambda_2
\Bigl[ \,( \alpha_{\lambda_2}\gamma_{\lambda_1} -
\alpha_{\lambda_1}\gamma_{\lambda_2}) - ({\rm c.c.}) \Bigr]\cr
&= d\lambda_1\wedge d\lambda_2
\Bigl\{ \,\Bigl[
{dz\over d\lambda_2} {d\over d\lambda_1} \Bigl( {b\over 2a^2}\Bigr)
-{dz\over d\lambda_1} {d\over d\lambda_2} \Bigl( {b\over 2a^2}\Bigr)
\Bigr]  - ({\rm c.c.}) \Bigr\}\; .\cr} \eqn\cdilm$$
Since $\lambda_1$ and $\lambda_2$ parameterize the complex
variable $z$ (the uniformizer on $R$), we can rewrite the above as
$$\eqalign{
\omega_{D}^{(2)} &= - dz\wedge \, d \Bigl( {b\over 2a^2}\Bigr)
+ d\bar z\wedge \, d \Bigl(\overline {b\over 2a^2}\Bigr)\cr
&= -dz\wedge d\bar z \Bigl[
\, {\partial\over \partial \bar z}  \Bigl( {b\over 2a^2}\Bigr)
+{\partial\over \partial z}  \Bigl( {\bar b\over 2\bar a^2}\Bigr) \Bigr]\, .
\cr} \eqn\twoform$$
This is the desired result for the two-form representing a dilaton insertion.
It is a completely general expression since it is constructed from the
most general family of local coordinates.

For the $\x$ state it is natural to define a one form,
$$\ket{\omega^{(1)}_\chi}\equiv
\sum_{i=1,2} d\lambda_i\,{\bf b}(v_{\lambda_i})\, \x \,\, .\eqn\oneformchi$$
Using \abstrivial, \coeff, and \binsert, we find
$$\ket{\omega_{\chi}^{(1)}} = \ket{0}\cdot\omega_{\chi}^{(1)} \;
,\eqn\chiform$$
where
$$\eqalign{
\omega_{\chi}^{(1)} &=\,-\,
 {1\over 2}\sum_{i=1,2} d\lambda_i\, (\beta_{\lambda_i}-
\ov\beta_{\lambda_i})\cr
&=  {1\over 2} \hbox{d} \ln \left( {a\over \bar a}\right)
- dz \left( {b\over 2a^2}\right) + d\bar z \left( {\bar b\over 2
\bar a^2}\right)\, .\cr}\eqn\oneform$$
Note that the dilaton two form and the $\x$ one-form are related
by
$$\omega_{D}^{(2)}= \,-\,d\omega_{\chi}^{(1)}\, , \eqn\ebgf$$
in agreement with \eketbrst, in view that $\d = -Q\x$.

We saw in sect.~2.1 that the dilaton state $\d$ is contained in
$\H$, but the state $\x$ is not.
Moreover, as we stated in sect.~2.5,
only states in $\H$ lead to insertions that are independent of
the phase of the local coordinates.
Therefore, the $\x$ insertion is expected to depend on the
phase of the local coordinate, but the $\d$ insertion is not.
We can now confirm this expectation. How does a
change in phase alter the family of local coordinates \famcoor ? Consider a
fixed point $p_0$ corresponding to some fixed values of
the parameters $(\lambda_1,\lambda_2)$,
and the local coordinate $w$ around $p_0$. For any point $p$ near $p_0$ a
change
in the phase of the local coordinate amounts to
$w(p)\rightarrow e^{-i\theta} w(p)$.
The value $z(p)$ must remain unchanged
when $w(p)$ changes, since the point $p$ does not move in the domain $R$,
and therefore its coordinate on the $R$-uniformizer $z$ should not change.
If we examine \famcoor\ we see that this requires
$a\to e^{i\theta}a$, $b\to e^{2i\theta}b$, etc. Under these transformations
we see immediately that
the dilaton form \twoform\ is invariant, while the
$\x$ one-form \oneform\ transforms nontrivially, as expected.

\noindent
\underbar{When local coordinates arise from metrics.} If the family of
local coordinates we have been using to compute the dilaton two form happens
to arise from a metric via the prescription reviewed in sect.~2.2, we can
simplify and interpret our result.
Using  \almfr\ in the expression for the two-form \twoform, we find
$$\omega_{D}^{(2)} = 2 dz\wedge d\bar z \, \partial \bar \partial \rho
= i\, R^{(2)} (\rho)\, ,\eqn\ctwoform$$
where, in the last step, we used the definition of curvature given
in \curvature.
For the state $\x$ we find
$$\omega_{\chi}^{(1)} =\,i\, [ \hbox{d}\theta_a -i(dz \partial \ln \rho-
d\bar z \bar\partial \ln\rho) ]\,\,, \eqn\chicurv$$
where $\theta_a$ is the phase of $a$. We mentioned earlier that the
prescription to find a family of local coordinates from a metric could
not determine a phase for $a$. If we have a chosen curve $\gamma$
along which we will integrate the $\x$ insertion we can naturally fix
$\theta_a = \theta_\gamma$, where $\theta_\gamma$ is the phase of the
tangent to the curve $\gamma$ (all in the $z$-uniformizer). The tangent
is defined uniquely by the orientation of the curve.
With this identification
we now find
$$\omega_{\chi}^{(1)}= i\,k \,.\eqn\geodchi$$
We have therefore shown that whenever local coordinates
arise from metrics $\d$ insertions compute bulk curvature
and $\x$ insertions compute geodesic curvature.

\chapter{Main Integral Formulae}

The dilaton two-form $\ket{\omega_D^{(2)}}$ and $\x$ one-form
$\ket{\omega_{\chi}^{(1)}}$ are used to integrate insertions of the
corresponding states over {\em fixed} Riemann surfaces.
This clearly does not suffice for our purposes.
Since we
deal with string amplitudes,
the insertions must be further integrated over {\em spaces} of
Riemann surfaces.
We will see that for the
$\K$-spaces and $\L$-spaces defined in sect.~3, the
insertion integrals over the
surfaces and their boundaries can be isolated from the integrals over the
rest of the data in these spaces (except possibly for the case of
surfaces without punctures, which we will not address in this paper.).
Such simplification was anticipated in Ref.[\distlernelson], where
it was observed that
the dilaton state $\d$ is annihilated by the product of any three antighost
oscillators of mode number greater than or equal to minus one.
The novel points of our discussion are that all our expressions are
well-defined since they never involve degenerate surfaces, we
recognize the importance of including the $\x$ state in the integral
statements, and the identities hold off-shell, namely other
states appearing in the correlators need not be physical.

The state $\x$ can also be
inserted into a space of surfaces by twist-sewing a three punctured
sphere carrying $\x$ on a special puncture. We will discuss how
the insertion of $\x$ depends on the choice of local coordinate
at the special puncture. The results of the present section pave
the way to the  later construction of the diffeomorphism that
establishes the dilaton theorem. They also illustrate how both
$\d$ and $\x$ insertions must act simultaneously to produce results that
only depend on Riemann surface data. This is important to understand
CFT deformations.

\section{Dilaton and $\x$ Insertions on $\K$-Spaces and $\L$-Spaces}

Let $\A$ be a $d$-dimensional space of surfaces of genus $g$
with $n>1$ punctures ($n\geq 3$ for $g=0$). We are interested in evaluating
the integrals
$$\eqalign{
f_D(\K\A) &\equiv  {1\over n!}\int_{\K\A} \bra{\Omega^{[d+2]g,n+1}\,}
\Psi\rangle_1\cdots\ket{\Psi}_{n}\ket{D}_{n+1}\,,\cr
f_\chi(\L\A) &\equiv  {1\over n!}\int_{\L\A} \bra{\Omega^{[d+1]g,n+1}\,}
\Psi\rangle_1\cdots\ket{\Psi}_{n}\x_{n+1}\,\, . \cr}\eqn\dillfun$$
If $\A$ is a single surface, the results of the previous
section for the dilaton two-form and the $\x$ one-form would suffice
to evaluate the integrals.
Recall that $\K$ and $\L$ insert states using some fixed but arbitrary
family of local coordinates.

We shall see that, for $\K\A$, the integration
of the dilaton two-form over the surfaces
can be carried out independently of the
integration over the space $\A$. For $\L\A$,
the integral of the $\x$ one-form over the boundaries of the surfaces
can be carried out independently of the
integration over the subspace $\A$.
The integrals over the surfaces and boundaries depend in general
on the surface,
but as we shall see, their sum {\it does not}, and
can therefore be factored out.
In addition, for local coordinates arising from a minimal area metric on the
surface the $\x$ integral vanishes, and the dilaton integral can be
factored out alone.

Let $(\xi_1,\ldots ,\xi_{d})$ denote a set of coordinates for the space $\A$.
It follows that
$\K\A$ is then parameterized by
$(\lambda_1,\lambda_2,\xi_1,\ldots ,
\xi_{d})$, where $\lambda_1,\lambda_2$ are two real parameters associated
to the position of the special
puncture that $\K$ brings in. A family of local coordinates at the special
puncture are defined by \famcoor, where now the coefficients depend on
all $d+2$ parameters.
Namely, the local coordinate for the dilaton insertion
depends on all the coordinates of the space $\A$, in addition to the
position of the puncture. Since the surfaces have $n+1$ punctures,
the Schiffer vectors must be $(n+1)$-tuples of vectors.
For the case of the $\x$ insertion over $\L\A$ similar remarks apply,
the only difference is that this space is $d+1$ dimensional.

Consider now the Schiffer vectors associated with the parameters
$\lambda_i$.
Since the $\K$-spaces and $\L$-spaces are defined by inserting an
additional puncture on each surface $\Sigma\in\A$, these $(n+1)$-tuples
cannot change any of the data on $\Sigma$. Therefore they need only be
supported at the special puncture,
$${\bf v}_{\lambda_i} = (0,\ldots ,0,v_{\lambda_i}^{(n+1)}) \,\quad
i=1,2\; ,\eqn\bsch$$
where the last entry is the Schiffer vector computed in section 2.4, which
moves the special puncture and changes its local coordinate.

Associated with the other $d$
parameters in the spaces $\K\A$ and $\L\A$, the
Schiffer vectors can be written in the form
$${\bf v}_{\xi_k} =  \wh{\bf v}_{\xi_k}+{\bf v}_{\xi_k}'\, ,
\quad k=1,\cdots, d\; ,\eqn\jko$$
where we have split the Schiffer vector into a vector $\wh{\bf v}_{\xi_k}$,
and a vector ${\bf v}_{\xi_k}'$ of the form
$$\wh{\bf v}_{\xi_k} =(v^{(1)}_{\xi_k},\ldots ,v_{\xi_k}^{(n)},0) \,,\quad
{\bf v}_{\xi_k}'=(0,\ldots ,0,
 v_{\xi_k}^{(n+1)})\, \, .\eqn\mofer$$
The vector $\wh{\bf v}_{\xi_k}$
is only supported on the original $n$ punctures of $\A$ and it
changes the coordinate $\xi_k$ of $\A$. Since a change in this data
must, in general, change
the data at the special puncture, this Schiffer vector must have
a component ${\bf v}_{\xi_p}'$ on the special puncture.
This component, however,
need only change the local coordinate at the special puncture. It follows that
$${\bf b} ({\bf v}_{\xi_p}') \ket{0}_{n+1} =0\,, \eqn\gue$$
since this antighost insertion only involves $b$-oscillators with mode
number greater than or equal to zero.

Having understood the nature of the Schiffer variations, we can now
use \cdefform\ to write the differential forms appearing in \dillfun\ as
(suppressing the string field insertions)
$$\eqalign{
\bra{\Omega^{[d+2]g,n+1}}D\rangle_{n+1} &=
\left( -2\pi i\right)^{2-n-3g} d\xi_1\wedge
\cdots\wedge d\xi_{d}\wedge d\lambda_1\wedge d\lambda_2\cr
&\,\,\quad\cdot \bra{\Sigma^{g,n+1}}\,{\bf b}(\bv_{\xi_1})\cdots
   {\bf b}(\bv_{\xi_{d}}){\bf b}(\bv_{\lambda_1})
\,{\bf b}(\bv_{\lambda_2}) \d_{n+1} \cr &\cr
\bra{\Omega^{[d+1]g,n+1}}\chi\rangle_{n+1} &=
\left( -2\pi i\right)^{2-n-3g} d\xi_1\wedge
\cdots\wedge d\xi_{d}\sum_{i=1,2} d\lambda_i\cr
&\,\,\quad\cdot \bra{\Sigma^{g,n+1}}\,{\bf b}(\bv_{\xi_1})\cdots
   {\bf b}(\bv_{\xi_{d}}){\bf b}(\bv_{\lambda_i})
 \x_{n+1} \; .\cr} \eqn\expform$$
We now recognize that the dilaton two-form defined in \dilmea\
and the $\x$ one-form defined in \oneformchi\ appear above.
Using \dilform\ and \chiform\ we can write the above equations as
$$\eqalign{
\bra{\Omega^{[d+2]g,n+1}}D\rangle_{n+1} &=
\left( -2\pi i\right)^{2-n-3g} \hskip-1pt d\xi_1\wedge
\cdot\cdot\wedge d\xi_{d}\wedge\omega_D^{(2)}
\, \bra{\Sigma^{g,n+1}}\,{\bf b}(\bv_{\xi_1})\cdot\cdot
   {\bf b}(\bv_{\xi_{d}})\ket{0}_{n+1} \cr &\cr
\bra{\Omega^{[d+1]g,n+1}}\chi\rangle_{n+1} &=
\left( -2\pi i\right)^{2-n-3g} \hskip-1pt d\xi_1\wedge
\cdot\cdot\wedge d\xi_{d}\wedge\omega_\chi^{(1)}
\, \bra{\Sigma^{g,n+1}}\,{\bf b}(\bv_{\xi_1})\cdot\cdot
   {\bf b}(\bv_{\xi_{d}})\ket{0}_{n+1}\; . \cr}\eqn\eformd$$
Note that the state at the $(n+1)$-th puncture is the vacuum state.
It cannot yet be brought all the way to the surface state $\bra{\Sigma}$ due
to the antighost insertions, that may have support on this puncture.
Nevertheless, using \jko\ and \gue,  we see that
$$\eqalign{\bra{\Sigma^{g,n+1}}{\bf b}(\bv_{\xi_1})\cdots
   {\bf b}(\bv_{\xi_{d}})\ket{0}_{n+1} &=
\bra{\Sigma^{g,n+1}}{\bf b}(\wh\bv_{\xi_1})\cdots
   {\bf b}(\wh\bv_{\xi_{d}})\ket{0}_{n+1} \cr
&= \bra{\Sigma^{g,n}}{\bf b}(\wh\bv_{\xi_1})\cdots
   {\bf b}(\wh\bv_{\xi_{d}})\; ,\cr}\eqn\simplant$$
where in the last step the vacuum state can go freely all the way to the
bra and delete the extra puncture. This enables us to rewrite the first of
Eqns.~\eformd\ as
$$\eqalign{
\bra{\Omega^{[d+2]g,n+1}}D\rangle_{n+1} &=
\left( -2\pi i\right)^{2-n-3g}\hskip-1pt d\xi_1\wedge
\cdot\cdot\wedge d\xi_{d}\wedge\omega_D^{(2)}
\cdot \bra{\Sigma^{g,n}}\,{\bf b}(\wh\bv_{\xi_1})\cdot\cdot
{\bf b}(\wh\bv_{\xi_{d}}) \cr
& = (-2\pi i)^{-1}\bra{\Omega^{[d]g,n}} \wedge \omega_D^{(2)}
\; , \cr}  \eqn\eform$$
and similarly,  the second one as
$$\bra{\Omega^{[d+1]g,n+1}}\chi\rangle_{n+1}
= (-2\pi i)^{-1}\bra{\Omega^{[d]g,n}} \wedge \omega_\chi^{(1)}\, .\eqn\eformp$$
The functionals of \dillfun\ then become
$$\eqalign{
f_D(\K\A) &=  {1\over n!}\int_\A
 \Big[\bra{\Omega^{[d]g,n}}\Psi\rangle_1\cdots\ket{\Psi}_{n}
\,\cdot {i\over 2\pi}
\hskip-2pt\int_{\Sigma - \cup D_i}\hskip-5pt\omega_D^{(2)}\Big]\cr
f_\chi(\L\A) &=  {1\over n!}\int_\A
 \Big[\bra{\Omega^{[d]g,n}}\Psi\rangle_1\cdots\ket{\Psi}_{n}
\,\cdot {i\over 2\pi}
\hskip-4pt\int_{\p(\Sigma - \cup D_i)}\hskip-7pt\omega_\chi^{(1)}\Big]\; .\cr}
\eqn\genform$$
We have thus succeeded in isolating the two dimensional
dilaton integral and the one-dimensional $\x$ integral
from the integral over the coordinates of the space $\A$.
If the local coordinates for the  insertions come from a metric on the
surfaces we can  use \ctwoform\ and \geodchi\ to write
$$\eqalign{f_D(\K\A) &= \,- \,\,{1\over n!} \int_\A
 \Big[\bra{\Omega^{[d]g,n}}\Psi\rangle_1\cdots\ket{\Psi}_{n}
\,\cdot {1\over 2\pi}\hskip-1pt
\int_{\Sigma - \cup D_i}\hskip-5pt R^{(2)}\Big]\cr
f_\chi(\L\A) &= \,-\,\,{1\over n!} \int_\A
 \Big[\bra{\Omega^{[d]g,n}}\Psi\rangle_1\cdots\ket{\Psi}_{n}
\,\cdot {1\over 2\pi}\hskip-3pt
\int_{\p(\Sigma - \cup D_i)}\hskip-7pt k\Big] \; .\cr}\eqn\genformm$$
Each of these integrals depends on the metric that we use for the surfaces
in $\A$. The sum, however, does not. Since the Euler number of the
bordered surface $\Sigma-\cup D_i$ is $(2-2g-n)$,
the Gauss-Bonnet theorem \gauss\ implies that
$$ f_D(\K\A) + f_\chi(\L\A) = (2g-2+n) f(\A)\, . \eqn\whoo$$
This is an important result.
It is independent of the metrics defined
on the surfaces and used to derive the local coordinates. It is therefore
well defined on spaces of Riemann surfaces.

A particular case of the above general result will
be useful to us. As we defined  in sect.~3.1, a space of surfaces $\V_i$ is one
in which the surfaces are
equipped with unit
disks that arise from minimal area metrics. In this
case the coordinate curves are geodesics and $f_\chi(\L\V_i)$ vanishes.
We then find that \whoo\ gives
$$f_D(\ov\K\V_i) = (2g_i-2+n_i)\,f(\V_i) \,  \, .\eqn\dilfunp$$
This result will be necessary in sect.~8 for the construction of the
diffeomorphism establishing the dilaton theorem.

\noindent
\underbar{Using the $\K^*$ operator} At the end of sect.~3.1 we mentioned
the possibility of using a $\K^*$ operator that would insert a puncture
with local coordinates induced from
a metric which is a deformation of the minimal area metric on the
interior of the surface minus its unit disks. The metric is left unchanged
in some neighborhood of the coordinate curves.
The purpose of this deformation is to produce metrics with smooth
curvature.
It is clear that for $\K^*$ insertions
$$f_D(\K^*\V_i) = (2g_i-2+n_i)\,f(\V_i) \,  \, ,\eqn\jufunp$$
since, as was the case with $\ov\K$ there is no geodesic curvature
contribution.
The following identities hold
$$\eqalign{
f_D\Bigl(\K^*\, (\,\{ \V_1 , \V_2\}\, )\, -\, \{ \K^*\V_1 , \V_2\}
 \,-\,\{ \V_1 , \, \K^* \V_2 \} \Bigr)\,=&\,\,0,\cr
f_D\Bigl(\K^* \, (\Delta \, \V_i) - \Delta\, (\K^* \V_i)\,\Bigr) =&\,\, 0
\,,\cr
[\,\partial\,,\,\K^*\,]\,\V_i\,+\,\{\, \V'_{0,3}\,,\V_i\,\} =&\,\, 0 \,.
\cr}\eqn\comnew$$
The first two identities are verified using \jufunp\ and
the first two equations in \hrehn. These equations hold strongly even
though the spaces of surfaces inside the big parentheses do not cancel
out, as they do for the $\ov\K$ operator (see Eqn.~\comdeltap).
The last identity holds because the metric is not changed in some
neighborhood of the coordinate curves.

The reader may recall that the recursive construction of $\B$ spaces
satisfying strong identities made use of $\ov\K$ and its properties.
As we see, $\K^*$ satisfies similar properties, as long as dilatons
are inserted on the special puncture. While we have not carried out
the analysis in detail, we  believe that a
discussion similar to that in sect.~3.2 would yield a sequence of
$\B$ spaces such that the recursion relations \improvet, with
$\ov\K$ replaced by $\K^*$, would hold upon inserting a dilaton
at the special puncture. This is all we would need to prove the dilaton theorem
using $\K^*$, as it will be clear in sect.~8.

\noindent
\section{Inserting $\x$ by Twist-Sewing}

The operation $\L$ acting on a space of surfaces $\A$ was defined to
add a puncture over the boundary $\p(\Sigma-\cup D_i)$ of every surface
$\Sigma\in \A$. The local coordinate at the puncture is taken to
arise from a metric on the surface.
The phase of that local coordinate can be fixed using
the tangent vector at the boundary, as explained at the end of
sect.~5. There is another way of inserting a puncture over the boundary
$\p(\Sigma-\cup D_i)$ of every surface
$\Sigma\in \A$. Let $\V_{0,3}^*$ be a standard two-punctured sphere,
with an additional special puncture.
The local coordinates around the first and second
punctures, located
at $z=0$ and at $z=\infty$, are given by
$$z_1 = z ,\quad\quad z_2 = 1/z \,, \eqn\descshp$$
where $z$ is a uniformizer. The local coordinate at the special puncture
situated at $z=1$ is given by
$$z_3 = a(z-1) + b(z-1)^2 + \cdots \, . \eqn\localcoord$$
The operation of twist-sewing $\V_{0,3}^*$ into a space of surfaces will
have the effect of inserting a puncture precisely on $\p(\Sigma -\cup D_i)$
for every surface $\Sigma$ in the space, as noticed
in \relko\ for the special case of $\ov \L$ and the particular sphere
$\V_{0,3}'$. The local coordinate at the
puncture will depend on the choice
\localcoord\ made in defining $\V_{0,3}^*$.
Note that the local coordinates
on $\V_{0,3}^*$ are well-defined even including phases.

The objective in the present section is to evaluate
$f_\chi \left( \{ \V_{0,3}^* , \A \}' \right)$
where $\A$ is some arbitrary space of surfaces.
We use
a prime on the antibracket to distinguish it from the standard antibracket
defined for symmetric spaces. Since the space $\V_{0,3}^*$ is not symmetric,
in $\{ \V_{0,3}^* , \A \}' $ we must choose one of the two ordinary
punctures of $\V_{0,3}^*$ to do the sewing. We will choose the first puncture,
and for the time being we will keep the choice of the local coordinate
at the special puncture open. Later on we will consider the simplification
that follows when the sphere $\V_{0,3}^*$ is required to be symmetric under
the exchange of the ordinary punctures at zero and infinity.

We can begin our evaluation of $f_\chi \left( \{ \V_{0,3}^* , \A \}'
\right)$
by making use of \hrehn\ and \takedelta
$$\eqalign{
f_\chi ( \{ \V_{0,3}^* , \A \}' ) &= (-)^{\A+1}\,  \{ f(\A)
\, ,f_\chi ( \V_{0,3}^*)\, \}' \cr
&= {(-)^{\A +1}\over (n-1)!}\int_{\A}
\bra{\Omega^{[k]g,n}_{1'\cdots n}}\Psi\rangle ^{n-1}\bra{V_{123}^*}
\Psi\rangle _2
\ket{\chi}_3\ket{\S_{11'}} \cr
&= \,{(-)^\A\over (n-1)!}\int_{\A}\bra{\Omega^{[k]g,n}}\Psi\rangle ^{n-1}
\bra{V_{123}^*}b_0^{-(1)}\ket{\Psi}_2\ket{\chi}_3\ket{R_{11'}}\; . \cr}
\eqn\chifunction$$
To proceed further we need to evaluate $\bra{V_{123}^*}b_0^{-(1)}$.
To this end, we use the Ward identity
$\bra{V_{123}^*}\sum_i\oint_{\C_i}{dz_i\over 2\pi i}b^{(i)}(z_i)
  v^{(i)}(z_i) = 0$,
where $v^{(i)}(z_i)$ is the restriction around the $i$-th puncture of
a holomorphic vector field $v(z)$ on the sphere minus the punctures.
An analogous relation holds for the antiholomorphic components.
Taking $v(z) = z$, we find
$$
v^{(1)} = z_1\,,\quad
v^{(2)} = -z_2\,,\quad
v^{(3)} = a + \left(1 + {2b\over a}\right)z_3 +\O(z_3^2) \; .\eqn\vectors$$
The Ward identity, together with its antiholomorphic counterpart, gives
$$\bra{V_{123}^*}\biggl[ b_0^{-(1)} - b_0^{-(2)} + (ab_{-1}^{(3)}-\bar a \bar
b_{-1}^{(3)})
  + \Bigl(1 + {b\over a} + {\bar b\over \bar a} \Bigr)
 b_0^{-(3)}+  \Bigl({b\over a} - {\bar b\over \bar a} \Bigr)
 b_0^{+(3)} + \cdots
  \biggr] = 0 \, ,\eqn\bidentity$$
where the dots indicate terms of the form $b_n^{(3)}$ and  $\bar b_n^{(3)}$
with $n\geq 1$.  Since $b_0^-$ annihilates the string field, and all
antighost oscillators $b_{n\geq-1},\bar b_{n\geq-1}$ except for $b_0^-$
annihilate $\x$, we have that
$$\bra{V_{123}^*}b_0^{-(1)} \ket{\Psi}_2 \x_3 \ket{R_{11'}} =
\Bigl(1 + {b\over a} + {\bar b\over \bar a} \Bigr)
\bra{V_{123}^*}0\rangle_3\ket{\Psi}_2  \ket{R_{11'}} =
\Bigl(1 + {b\over a} +{\bar b\over \bar a}\Bigr)\ket{\Psi}_{1'}\,.\eqn\frd$$
Back in \chifunction\ we then find
$$f_\chi ( \{ \V_{0,3}^* , \A \}' )  = (-)^{\A}\,n\,
\Bigl(1 + {b\over a} + {\bar b\over \bar a} \Bigr) f(\A)\,. \eqn\ction$$
This is the equation we were after. It shows that line integration
of a $\x$ insertion along the coordinate curves of surfaces amounts to
multiplication
by a constant determined by the geometry of a three string vertex
that fixes the way $\x$ is inserted.

The three string vertex $\V_{0,3}^*$ we are going to use in our
applications
will be the vertex $\V'_{0,3}$ discussed earlier, and fixed to be symmetric
under
the exchange of punctures one and two. That means that $z_3(z) = -z_3(1/z)$.
It is straightforward
to show that this requirement on
\localcoord\ leads to  $b=-a/2$. It then follows that for the symmetric vertex
$\V'_{0,3}$ the line integrals of $\x$ gives zero, namely
$f_\chi ( \{ \V'_{0,3} , \A \}' ) = 0$. Since $\V'_{0,3}$ is symmetric
we can use the standard antibracket
$$f_\chi ( \{ \V'_{0,3} , \A \} ) = 0 \, .\eqn\zero$$
This result is intuitively plausible
in the light that $\x$ insertions
compute geodesic curvature. Assume a $\x$-insertion along the
coordinate curve of some puncture is done via $\L$ using some metric on
the surface. We expect
that such an insertion can be reproduced by twist sewing with a fixed
sphere $\V_{0,3}^*$
if the metric in some neighborhood of the coordinate curve
is invariant under rigid rotations
of the local coordinate at the puncture.
If the coordinate curve is a geodesic
of the metric we would expect that the
sphere that reproduces the $\L$ insertion would have a symmetric coordinate at
the special puncture.
In this situation both the $\L$ insertion and
the twist-sewing insertion would give zero.

\chapter{CFT Deformations and String Backgrounds}

In this section we first use the
dilaton to generate a CFT deformation. This deformation
cannot be obtained in the standard way since the dilaton is not primary.
It must involve the $\x$ state. We then show
that this deformation is {\em trivial}.
The second part is
an attempt to give a sensible definition of a string background,
and to argue that the dilaton deforms it in a {\em nontrivial} way.

\section{The CFT Deformation associated with the Dilaton}

The main question we want to address here is whether or not
the dilaton state is an exactly marginal state. In other words
we would like to know if this state generates an exact CFT deformation.
When physicists discuss exactly marginal states of a matter conformal
field theory they have in mind primary states $\ket{\O_M}$ of dimension
$(1,1)$.
When including the ghost CFT, one typically considers
the dimension $(0,0)$ primary
states $c_1\bar c_1\ket{\O_M}$.
Since they are primary, these states lead to well defined two-forms on
the surface, namely two-forms that are independent of the local
coordinates used to insert the states. The insertions can therefore be
integrated unambiguously over Riemann surfaces.
This is essential since Riemann surfaces do not come equipped with canonical
families of local coordinates.
As we have seen in sect.~2.1,
the dimension $(0,0)$ dilaton state $\d$ is not of the
standard form and is not primary.
 As we will see, one cannot describe the deformation of a CFT
just using $\d$, one must use the pair
$( \,\d , \x )$ to be able
to write something well defined on Riemann surfaces.
After explaining how the deformation can be written,
we will see that due to the ghost number anomaly it turns out to be trivial.

Now let us try to write explicitly a deformation
induced by the dilaton.
Given a surface state, a canonical deformation can be obtained by integration
of
the marginal operator over the surface minus its unit
disks [\campbell,\ranganathan,\rangasonodazw]
$$\delta\, \bra{\,\Sigma\, } = -\,{\epsilon\over\pi} \hskip-6pt
\int_{\Sigma - \cup D_i}\hskip-6pt d^2 z\bra{\,\Sigma;z\,}
\O_M\rangle\,.\eqn\xen$$
Writing this deformation in the language of forms
we find
$$\delta\, \bra{\,\Omega_\Sigma^{[0]g,n}\, }
= \epsilon \int_{\K \Sigma }
\bra{\Omega^{[2]g,n+1}}c\bar c\O_M\rangle_{(n+1)} \,,\eqn\xenf$$
where the zero-form on the left hand side is just the surface state
up to a constant factor implicit in \cdefform.\foot{The constants in \xenf\
were worked out in Ref.[\senzwiebachtwo] below Eqn.~(3.25).}
Guided by the above equation, we attempt to define the dilaton
deformation by
$$\delta\, \bra{\,\Omega_\Sigma^{[0]g,n}\, }  \, =
\, \epsilon \int_{\K \Sigma }
\bra{\Omega^{[2]g,n+1}}D\rangle_{(n+1)} \; .\eqn\dildef$$
The obvious difficulty with this expression is that since the dilaton is
not primary, the integral is not well defined. If we put a metric on
the surface and use it to obtain local normal-coordinates, the integral
depends on the metric. Indeed, using \eform\ and \ctwoform\ we see that
$$\delta\, \bra{\,\Omega_\Sigma^{[0]g,n}\, }  \,
=  - \, \epsilon \bra{\Omega^{[0]g,n}}\,\cdot\,
{1\over 2\pi}\hskip-3pt\int_{ \Sigma-\cup D_i }\hskip-6pt R^{(2)}\;
.\eqn\dief$$
Given that the integral depends on the chosen metric, this is not
a well-defined deformation on a Riemann surface. The cure is clear from
our earlier developments, we must include in the deformation a contribution
from $\x$ that will give geodesic curvature.
$$\delta\,\bra{\,\Omega_\Sigma^{[0]g,n}\, } \, =  \, \epsilon
\int_{\K \Sigma }
\bra{\Omega^{[2]g,n+1}}D\rangle_{(n+1)} \, + \epsilon
\int_{\partial(\K \Sigma)}
\bra{\Omega^{[1]g,n+1}}\chi\rangle_{(n+1)}\, .\eqn\dildef$$
This will give
$$\delta\, \bra{\,\Omega_\Sigma^{[0]g,n}\, }  \,
=  - \, \epsilon \bra{\Omega^{[0]g,n}}\,\cdot\,\Bigl(
{1\over 2\pi}\hskip-3pt\int_{ \Sigma-\cup D_i }\hskip-6pt R^{(2)}
+\,{1\over 2\pi}\hskip-6pt\int_{ \p(\Sigma-\cup D_i) }\hskip-8pt k \,
\Bigr) \; ,\eqn\dieff$$
and this time the sum of the integrals gives the Euler number
for any choice of the conformal metric. Thus the above is well defined
on Riemann surfaces.
In terms of the surface states we then have
$$\delta\, \bra{\Sigma_{g,n}} \,= - \,\,\epsilon\, (2-2g-n)\,
\bra{\Sigma_{g,n}} .\eqn\ddeff$$
The deformation induced by the dilaton
amounts to a scaling
of the surface states of the conformal theory.
Since the conformal theory is taken to have
zero central charge, the operator formalism surface states have
well defined normalization and it is possible to discuss the change
of normalization.

We can now verify explicitly that \ddeff\ is indeed a CFT
deformation. In a CFT, surface states form a representation of the operation
of sewing. Given two surfaces $\Sigma_1$ and $\Sigma_2$ which
can be sewn together to form the surface $\Sigma_1\cup\Sigma_2$,
the corresponding surface states must satisfy
$\bra{\Sigma_1\cup\Sigma_2}=\bra{\Sigma_1}\bra{\Sigma_2} R_{12}\rangle$,
where the ket $\ket{R_{12}}$
is the inverse of the bra $\bra{R_{12}}$
representing  the canonical two-punctured sphere.
It follows from \ddeff\ that $\delta\,  \bra{R_{12}} = 0$, and therefore
the inverse relation requires that $\delta\,  \ket{R_{12}} = 0$.
We readily verify that we have a CFT deformation:
$$\eqalign{
&(\bra{\Sigma_1} +  \delta \bra {\Sigma_1})\,
(\bra{\Sigma_2} +  \delta \bra {\Sigma_2})\,
\ket{R_{12}}  \cr
&=\bra{\Sigma_1}\bra{\Sigma_2} R_{12}\rangle \Bigl( 1- \epsilon \,[
2-2g_1 -n_1 + 2 - 2g_2 -n_2]\, \Bigr)\,\,,\cr
&= \bra{\Sigma_1\cup\Sigma_2}\Bigl( 1- \epsilon \,[
2-2(g_1 + g_2)-(n_1+n_2-2) ]\, \Bigr)\cr
&=\bra{\Sigma_1\cup\Sigma_2} + \delta\, \bra{\Sigma_1\cup\Sigma_2}
\,. \cr}\eqn\vdef$$
Indeed we see that sewing the deformed states gives precisely the
deformation of the sewn state. The question is whether the deformed CFT is new,
or is just the old one in a different basis.
To answer this question, consider the most general change of basis in
the CFT
$$\ket{\Psi}\to \ket{\Psi} - \epsilon \,A\ket{\Psi}\, ,\eqn\basisch$$
where $A$ is an arbitrary but fixed matrix of constants in the CFT state space.
Under this change of basis the surface states change as
$$\bra{\Sigma_{g,n}} \to \bra{\Sigma_{g,n}}+ \epsilon \sum_{i=1}^n
\bra{\Sigma_{g,n}}A^{(i)}\, ,\eqn\chast$$
so that the correlators are preserved.
In here the superscript $i$ denotes
the label of the state space that the matrix $A$ acts on. The transformation
in \chast\ is a
CFT deformation since, as one readily verifies, it preserves sewing. It is
a trivial deformation because it arises from a change of basis.
If the deformation in \ddeff\ is trivial it must follow that for some choice of
$A$ one must have that
$$\bra{\Sigma_{g,n}} \Bigl( 2-2g-n + \sum_{i=1}^n
A^{(i)}\Bigr) = 0\, , \eqn\isittriv$$
for {\it any}  choice of surface $\Sigma_{g,n}$.
If we take $A= {\bf 1}$, we see that we can eliminate the $n$ term out
of the deformation. In fact, taking $A$ to be an arbitrary constant times
the identity, we see that
$ \delta\,\bra{\Sigma_{g,n}}= -\e\, (2-2g-
\beta n)\, \bra{\Sigma_{g,n}}$ is a CFT deformation
for any value of the constant $\beta$. This shows that the $n$ term in
\ddeff\ is completely ambiguous, and thus it is no surprise that
Distler and Nelson found such ambiguities (see the discussion of Eqn.~\distnel\
in the introduction).

This is not the end of the story, however. The term $(2-2g)$ in the deformation
is also trivial. This is due to the ghost number anomaly. Recall
that the ghost current operator is not primary, and as a consequence the
ghost number operator, ordinarily written as
$$ G = \sum_n :c_nb_{-n}  + \bar c_n \bar b_{-n} :   \; ,\eqn\ghostn$$
where the normal ordering is with respect to the $SL(2,C)$ vacuum, exhibits
anomalous behavior.
In particular, acting on surface states it gives
$$\bra{\Sigma_{g,n}\,} \sum_{i=1}^n G^{(i)} = (6-6g)\,\bra{\Sigma_{g,n}\,}\,.
\eqn\ghanomali$$
It now follows that \isittriv\ is satisfied by choosing $A= {\bf 1} -G/3$.
This completes our proof that the dilaton deformation of a CFT is
trivial. \foot{In Ref.[\bankssen]
the $\sigma$-model dilaton was
coupled to the ghost current rather than
to the curvature [\tseytlin]. A better understanding
of the relation between the two approaches might yield some insight
into the fact that zero momentum dilatons do not deform the CFT.}

\section{Dilatons Deform String Backgrounds}

There is more to a closed string background than a conformal field theory.
In a conformal theory the natural objects are the surface states, which
encode the information about the correlators on a
{\em fixed}~ Riemann surface.
The set of objects necessary to build a string field theory includes
integrated correlators, or forms over moduli spaces of Riemann surfaces.
While we could find a similarity transformation
that canceled out the dilaton deformation of the surface states (the
zero-forms), the same transformation cannot cancel out the
dilaton deformation of the general forms. This is simple to see.
Since the similarity operator was generated by ghost number,
the antighost insertions necessary to build the general forms
will change the counting that made the ghost number work for the
zero-forms.

We will not try to propose a fully axiomatic definition of a
string background. The
following definition is just an attempt to capture the
basic idea.\foot{There have been
some attempts to systematize the idea of a
string background. See, for example [\sentalk], where the
algebraic structure of classical string field theory was discussed, or the
discussion based on homotopy Lie algebras and their extensions [\zwiebachlong].
More recent works, by Kimura, Stasheff, and Voronov, [\voronov,\voronovkimura]
also propose a definition of a (tree-level) string background that
arises from a conformal theory. We believe that a physically motivated
definition of a string background should not assume a
conformal theory, and must incorporate loops.
The definition we propose, in a somewhat
different form, was suggested by E. Witten as a tool to
attempt a definition of ``theory space''.}

\noindent
\underbar{String Background}
{}~Let $\H$ be a Batalin-Vilkovisky vector space
equipped with a function $S_{0,2}$, and let $\wh\P$ denote the space of
subspaces of $\cup_{g,n} \wh\P_{g,n}$. A string background furnishes a map
from $\wh\P$ to the space of
functions on $\H$.
In other words, given a space of surfaces
$\A_{g,n}\subset \wh\P_{g,n}$ we are given a function $f(\A_{g,n})$
in $\H$. This map must satisfy the consistency conditions
$$\eqalign{
f\bigl( \Delta \A ) &= -\Delta  f(\A)  \cr
f \bigl( \{ \A , \B \} \bigr) &= - \{\, f(\A) , f(\B) \} \cr
 f(\p\A) &= -\{ S_{0,2} , f(\A)\}\; .\cr}\eqn\hrehm$$
This definition includes more than what we really need to construct
a string field theory.
A possible variant could be to demand the above conditions only for
spaces $\A_{g,n}$ of the same dimension as $\M_{g,n}$.
If we find an $f$ satisfying the above conditions, we have a string background.
An obvious $f$ was given in \fgeomty.

We can now think of deformations. The obvious one to try is
$$f(\A_{g,n}) \to f(\A_{g,n}) \Bigl( 1 - \e \,[2-2g-n]\Bigr)\,,
\eqn\defcback$$
which would follow by integrating, for each surface in the space,
a dilaton over the surface minus its unit disks, and $\x$ over the
boundaries of the disks. As we saw in sect.~6, the integral over the
space $\A$ does not interfere with the integrals over the surfaces.
In fact the above deformation is just the same as letting $f(\A)\to
f(\A) + \e (f_D(\K \A)+  f_\chi (\L \A)) $. In addition,
we do not deform $S_{0,2}$.
It is straightforward
to verify that \defcback\ is indeed a  deformation of $f$ preserving
Eqns.~\hrehm. It is therefore a deformation of a
string background.

It follows from \fgeomty\ that at the level of forms, the deformation
postulated in \defcback\ arises from the deformation
$$\bra{\Omega^{[d]g,n}} \to  \bra{\Omega^{[d]g,n}}
\Bigl( 1 - \e \,[2-2g-n]\Bigr)\,,\eqn\atlof$$
where $d$, the degree of the form,  is equal to the number of
antighost insertions.
Acting on these forms the ghost number operator gives
$$\bra{\Omega^{[d]g,n}} \sum_{i=1}^n G^{(i)} =   \bra{\Omega^{[d]g,n}}
\Bigl( 6-6g +d\Bigr)\,,\eqn\atloff$$
where use was made of \ghanomali\ after commuting the ghost number operators
through the antighost insertions (recall $[G, b] =-b$).
It is clear from \atlof\ and \atloff\
that a similarity transformation induced by the
ghost number operator
will not be able
to reproduce the deformation of the
various forms.\foot{More precisely, the similarity
transformation must be a canonical transformation, that is, it should
preserve the symplectic structure in $\H$. The ghost number
generated canonical transformation reads
$\delta\ket{\Psi} = \e(G-5/2) \ket{\Psi}$. Note that the similarity
transformation $\delta\ket{\Psi} = \e\ket{\Psi}$ is not canonical. When
doing string field theory this redefinition changes the observables of the
theory unless the symplectic form is changed accordingly (see sect.~6.1.3
of Ref.[\senzwiebachtwo]).}
We believe there is no
similarity transformation that can reproduce this deformation, namely we
believe that the dilaton truly deforms a string background. The existence
of a similarity transformation generating this deformation would imply that
there is a homogeneous string field transformation that changes the
coupling constant of string theory. This would make no sense, since it would
mean that the string coupling constant is not an observable. The fact that
a {\it shift} of the string field is necessary (along with a linear and higher
order transformations) to change the string field coupling constant,
is essential to have a change in the physics of the
theory. Indeed, at the
level of low energy Lagrangian, it is well known we must shift the
dilaton to change the string field coupling constant.

\chapter{The Diffeomorphism Establishing the Dilaton Theorem}

In sect.~4 we gave the precise statement of the dilaton theorem. We concluded
that we had to find a Hamiltonian function ${\bf U}_D$
satisfying \maine
$$\eqalign{
\{ S , {\bf U}_D \} +
\hbar \Delta {\bf U}_D &= a\hbar \k \, {d\over d\k}
\left(  S_{1,0} + {1\over 2}  \ln \rho\right) \cr
&\quad + a \sum_{g,n}(2g\hskip-1pt -\hskip-1pt 2+n)\,\hbar^g\k^{2g-2+n}
\,f(\V_{g,n}) \, .\cr} \eqn\mainee$$
On the right hand side of this equation we included  a term related
to the one-loop free energy. Since the genus one contribution to the
free energy $F_{g=1}$ is usually thought to be coupling constant independent,
this term is a surprise. We will see that it may be necessary.
We also saw in sect.~4 that the
Hamiltonian function ${\bf U}_D$ is expected to be of the form
(Eqn.~\truncgt)
$$ {\bf U}_D = {1\over \k} \,  \, {\bf U}_{D(0,2)} + f_\chi
(\underline\V_{0,3}) +\cdots
\, . \eqn\diliff$$
The object of the present section is to give the missing terms in
the above equation and thus find the
diffeomorphism that establishes the dilaton theorem.
We claim that the
complete diffeomorphism is given by
$$ {\bf U}_D = {1\over \k} \, \Bigl[ \, {\bf U}_{D(0,2)} + f_\chi
(\k\underline\V_{0,3}) -
f_D (\B_>) + f_\chi ( \{ \k\B_{0,3}\, , \V \} ) \Bigr]\, , \eqn\dildiff$$
where the subscript $>$ denotes that the $(0,3)$ space is not
included (see Eqn.~\introdnew).
The third term is the same term that appears in
the diffeomorphism implementing background independence except for the
exclusion of $\B_{0,3}$. The last term is new, and was found by
trial and error. Here the $\x$ state
is inserted on the special puncture of the three-punctured spheres
defining the space $\B_{0,3}$.
Even though $\x$ is not in $\H$, this term is well-defined
since we can define unambiguously a
local coordinate with a phase on the special puncture of the spheres
defining $\B_{0,3}$.
It should be noted that the
last two terms in \dildiff\ give contributions to the Hamiltonian that begin
as cubic terms in the string field. They complete the quadratic Hamiltonian
that was anticipated by the analysis of gauge transformations.

We now verify that the Hamiltonian ${\bf U}_D$ does indeed satisfy
\mainee.
The first term on the left hand side of \mainee\ is given by
$$\eqalign{
 \{ S , {\bf U}_D \} = &\quad {1\over \k} \,\{ S , {\bf U}_{D(0,2)} \}  \cr
& +\{ S , f_\chi ( \underline\V_{0,3}) \} \cr
& -{1\over \k} \, \{ S ,  f_D ( \B_> )\} \cr
&  + \{ \, S ,\,\,  f_\chi ( \{ \B_{0,3} , \V \})\,\, \}\,. \cr}\eqn\asic$$
Using \excellent\ and \beexcellent, we immediately find
$$\eqalign{
 \{ S , {\bf U}_D \} &= {1\over \k} \,f_D ( \underline\V ) \cr
& - f_D ( \underline\V_{0,3})  - f_\chi ( \{ \V , \underline \V_{0,3} \} )  \cr
&+ {1\over \k} \,  f_D \Bigl(\partial\B_> +
          \{ \V , \B_>\} \Bigr) \cr
&  - f_D (\{ \B_{0,3} , \V \})
 -  f_\chi \Bigl( \partial \{ \B_{0,3} , \V \} +
\{ \V , \{ \B_{0,3}\,, \V \}    \} \Bigr) \,, \cr}\eqn\assic$$
where each row in \assic\ is equal to the corresponding row in \asic\ .
Collecting terms and using the Jacobi identity \jcbdntty\ in the fourth
row gives
$$\eqalign{
 \{ S , {\bf U}_D \} &= {1\over \k} \,f_D \Bigl(
 \partial\B_>  + \{ \V , \B \} + \underline\V_> \Bigr) \cr
&  -  f_\chi \Bigl( \{ \underline\V_{0,3} + \p\B_{0,3} , \V \} +
\{ \B_{0,3} , \p\V + {1\over 2}\{ \V , \V \}\}\Bigr) \,.\cr}\eqn\assic$$
Using the geometrical recursion relations \sumrecursions,
the  relation $\p\B_{0,3} = \V'_{0,3}- \underline\V_{0,3}$, and
the vanishing of $f_\chi ( \{ \V_{0,3}' , \V \})$ (Eqn.~\zero), we obtain
$$\{ S , {\bf U}_D \} = {1\over \k} \,f_D \Bigl(
 \partial\B_>  + \{ \V , \B \} + \underline\V_> \Bigr)
+  f_\chi \Bigl(
\{ \B_{0,3} , \hbar\Delta\V\}\Bigr)\, .\eqn\assicx$$
This completes the calculation of the first term on the
left hand side of \mainee. The second term
is easily calculated using the first equation in \hrehn, and the second
equation in \edeltasquare\
$$\hbar \Delta {\bf U}_D = -\,\hbar f_\chi\, (\Delta \underline\V_{0,3})
+ {1\over \k}\, f_D
(\hbar\Delta\B_>) -\, f_\chi\,(\{ \B_{0,3}\,, \hbar \Delta \V \} )\,.\eqn\nbv$$
The last two equations add up to give
$$\hbar \Delta {\bf U} _D+ \{ S , {\bf U}_D \} = {1\over \k} \,f_D \Bigl(
 \partial \B_>  + \{ \V , \B \} + \hbar \Delta \B_> +
\underline\V_> \Bigr)
 -\,  f_\chi (\hbar \Delta \underline\V_{0,3})\,\,. \eqn\aseeicc$$
Using the recursion relation \improvett\ for the $\B$ spaces we finally find
$$\hbar \Delta {\bf U}_D+ \{ S , {\bf U}_D \} =  \,f_D (\ov\K\V) +
\hbar\, [ \, f_D (\underline\V_{1,1})
-\,  f_\chi (\Delta \underline\V_{0,3})\, ] .\eqn\aseeic$$
We are now ready to verify \mainee.
In order to do this, the first term in the right hand side
is separated into terms
having to do with ordinary string vertices, and terms having to do
with vacuum vertices (for $g\geq 2$)
$$ f_D (\,\ov\K \V\, ) = \sum_{g,n\geq 1} \hbar^g
\k^{2g-2+n}
\,f_D ( \ov \K \V_{g,n}) + \sum_{g\geq 2} \hbar^g
\k^{2g-2}\,f_D ( \ov \K \V_{g,0})\; . \eqn\jackg$$
Using \dilfunp\ for the $n\geq 1$ terms, we find
$$f_D (\,\ov\K \V\, )
= \sum_{g,n\geq 1} \hbar^g \k^{2g-2+n}\,
(2g-2+n) \,f (\V_{g,n}) + \sum_{g\geq 2} \hbar^g
\k^{2g-2}\,f_D ( \ov \K \V_{g,0})\; .  \eqn\jackgian$$
Comparing with \mainee, we see that all the string field dependent
terms of the equation can now be made to match by choosing
$a=+1$.
This shows that shifting the dilaton amounts to a shift in the coupling
constant in all the string field dependent terms of the string action.
This establishes
the dilaton theorem as far as all off-shell amplitudes is concerned.

Comparing the string field independent terms in \mainee\
we are led to the requirements
$$ f_D (\underline\V_{1,1}) -\,f_\chi ( \Delta \underline\V_{0,3}) =
\k \, {d\over d\k} \left(  S_{1,0} + {1\over 2}  \ln \rho\right)
\,\,,\eqn\condone$$
at genus one, and
$$f_D(\ov\K\V_{g,0}) = (2g-2)\, f(\V_{g,0} )\,\,,\eqn\condtwoalso$$
for $g\geq 2$.
For the latter equation, \dilfunp\ is not directly applicable
since the surfaces in the moduli space $\V_{g,0}$ have no punctures.
One might  compute
$f_D(\ov\K\V_{g,0})$ by adding an auxiliary puncture, and then following the
procedure of sect.~3.2.
The genus one case is also unclear.
Since $\ket{D}=-Q\ket{\chi}$,
one might be tempted to use \importantr\ to claim that
$$f_D(\underline\V_{1,1}) = - f_{\chi}(\partial\underline\V_{1,1})\, ,
\eqn\boundary$$
and the recursion relation
$$\partial\V_{1,1} = -\hbar\Delta\V_{0,3} \, ,\eqn\vertex$$
to conclude that the left hand side of \condone\ vanishes.
However each step in this analysis is problematic.
\boundary\ only holds if there exists a section in
$\P_{1,1}$ over $\V_{1,1}$,
and it is not clear that one does exist. In
addition,  \vertex\ is only known to hold in $\wh\P_{1,1}$ and
may not hold in $\P_{1,1}$. We leave open
the issue of vacuum graphs and the dilaton.

\ack  We wish to thank K. Ranganathan, A. Sen, and H. Sonoda for
their constructive comments during the course of this research.
One of us (O.B.) wishes to thank the Center for Theoretical Physics
and the Department of Physics at M.I.T. where part of this work was
done.

\refout

\end